\documentclass{article}
\usepackage[a4paper,left=2.4cm,right=2.4cm,bottom=2.4cm,top=2.4cm]{geometry}
\usepackage{soul} 
\usepackage{amsmath,amssymb} 
\usepackage{xcolor} 
\usepackage{graphicx}
\usepackage{rotating}
\usepackage{subfig}
\usepackage{pdfpages}
\usepackage{authblk}

\begin{document}

\title{A neutral comparison of statistical methods for time-to-event analyses under non-proportional hazards}

\author[1]{Florian Klinglm\"uller}
\author[1]{Tobias Fellinger}
\author[2]{Franz K\"onig} 
\author[3]{Tim Friede}
\author[4]{Andrew C. Hooker} 
\author[2]{Harald Heinzl}
\author[2]{Martina Mittlb\"ock} 
\author[2,3]{Jonas Brugger}
\author[3]{Maximilian Bardo} 
\author[3]{Cynthia Huber}
\author[3,5]{Norbert Benda} 
\author[2,*]{Martin Posch}
\author[2]{Robin Ristl}

\affil[1]{Austrian Agency for Health and Food Safety, Vienna, Austria}
\affil[2]{Medical University of Vienna, Center for Medical Data Science, Vienna, Austria}
\affil[3]{University Medical Center Göttingen, Department of Medical Statistics, Göttingen, Germany}
\affil[4]{Dept. of Pharmacy, Uppsala University, Uppsala, Sweden}
\affil[5]{Federal Institute for Drugs and Medical Devices (BfArM), Research Division, Bonn, Germany}

\affil[*]{Corresponding author, e-mail: martin.posch@meduniwien.ac.at}

\maketitle

\begin{abstract}
\noindent While well-established methods for time-to-event data are available when the proportional hazards assumption holds, there is no consensus on the best inferential approach under non-proportional hazards (NPH). However, a wide range of parametric and non-parametric methods for testing and estimation in this scenario have been proposed.
To provide recommendations on the statistical analysis of clinical trials where non proportional hazards are expected, we conducted a comprehensive simulation study under different scenarios of non-proportional hazards, including delayed onset of treatment effect, crossing hazard curves, subgroups with different treatment effect and changing hazards after disease progression. We assessed type I error rate control, power and confidence interval coverage, where applicable, for a wide range of methods including weighted log-rank tests, the MaxCombo test, summary measures such as the restricted mean survival time (RMST), average hazard ratios, and milestone survival probabilities as well as accelerated failure time regression models. 
We found a trade-off between interpretability and power when choosing an analysis strategy under NPH scenarios. While analysis methods based on weighted logrank tests typically were favorable in terms of power, they do not provide an easily interpretable treatment effect estimate. Also, depending on the weight function, they test a narrow null hypothesis of equal hazard functions and rejection of this null hypothesis may not allow for a direct conclusion of treatment benefit in terms of the survival function. 
In contrast, non-parametric procedures based on well interpretable measures as the RMST difference had lower power in most scenarios. Model based methods based on specific survival distributions had larger power, however often gave biased estimates and lower than nominal confidence interval coverage. The application of the studied methods is illustrated in a case study with reconstructed data from a phase III oncologic trial.
\end{abstract}

\newpage

\section{Introduction}
In randomized controlled clinical trials where time-to-event outcomes are the primary endpoint, current standard approaches to compare survival functions are the logrank test and reporting hazard ratios estimated by Cox proportional hazards models. Under the assumption of proportional hazards, the logrank test is the most powerful rank-invariant test for the comparison of time-to-event outcomes between treatment groups \cite{peto1972asymptotically}, and the Cox model hazard ratio provides a well interpretably estimate for the treatment effect.

However, in many settings, the proportional hazard assumption is unlikely to hold and analysis methods suitable for non-proportional hazards (NPH) are warranted.
Exemplary scenarios that can lead to NPH are a delayed onset of the treatment effect (as it has been described for immuno-oncology drugs, \cite{anagnostou2017immuno}), heterogeneous treatment effects between patient subgroups, or due to intercurrent events \cite{shen2023nonproportional}, or heterogeneity in patient frailty in general \cite{aalen1994effects}.

Under NPH, the power of the logrank test and asymptotically equivalent tests derived from the Cox model may be low and the interpretability of hazard ratio estimates is limited.
First, the value of the estimated hazard ratio depends on the study-specific censoring pattern \cite{xu2000estimating}. 
The corresponding estimated parameter is some weighted average of the hazard ratio function, whose weights depend on the censoring distribution. Therefore, this parameter has no direct interpretation.
Second, differences at specific time-points between the hazard functions of a treatment and a control group do not necessarily allow for a causal interpretation of a treatment effect at that time point \cite{hernan2010hazards}. 
Rather, the hazard ratio at a specific time point compares populations of trial participants that did not experience the event of interest until the specific time point. These are, however, selected populations that may differ systematically (in contrast to the randomised treatment groups). Therefore, variations in the hazard ratio over time can be due to this selection mechanism rather than due to a time dependence of the treatment effect \cite{hernan2010hazards,martinussen2020subtleties,aalen2015does,posch2022testing}. 
In contrast, differences or ratios between landmark survival probabilities, median survival times (or other quantiles) and restricted mean survival times \cite{bartlett2020hazards}, as well as measures based on accelerated failure time models \cite{aalen2015does} have been suggested as better interpretable effect size measures. 

Furthermore, when interpreting hypothesis tests to test for differences of survival functions under NPH, the null hypothesis needs to be carefully defined. The logrank test, e.g., is designed to test the null hypothesis of equal hazard functions or, equivalently, equal survival functions. Under proportional hazards, this is the only null hypothesis of interest, as any deviation can be interpreted as superiority for one of the two treatment groups. Under NPH, 
a broader definition of the null hypotheses may be required to account, for example, for scenarios where hazard functions cross while the cumulative hazard and hence the survival probability is always favorable under control compared to treatment \cite{magirr2019modestly}. Especially,  hypotheses tests should control the type 1 error rate under the null hypothesis that the survival function under treatment is lower or equal than under control. 

In addition, the considered tests and estimators should be robust with regard to other study characteristics as the censoring distribution and different lengths of follow-up.
Finally, the power of the procedures is of concern as well as the related implications for trial planning, such as the required sample size, number of events and (minimum) follow-up time. 

To address the shortcomings of the standard analysis approaches in the presence of NPH, a range of alternative procedures has been proposed \cite{bardo2023methods}. For hypothesis testing, weighted log-rank tests have been considered that have a larger power than the unweighted test, if more weight is allocated to time points where the effect is larger \cite{fleming1981class,zucker1990weighted,jimenez2019properties,liu2018weighted}.
Extensions of these tests that simultaneously take multiple weighting vectors into account, can increase the robustness of these tests if there is uncertainty with regard to the expected time pattern of the effect \cite{lee2007versatility,fleming2011counting,lin2020alternative,ristl2021delayed}. 

Other approaches are based on specific summary measures of the difference in survival curves, such as the restricted mean survival time (RMST) \cite{royston2011use}, average hazard ratios \cite{rauch2018average}, difference in median survival time or other quantiles of the survival function, and the analysis of x-year survival probabilities (milestone analyses).
In addition, parametric approaches have been proposed that are based on specific distribution functions (as, e.g., Weibull distributions) and estimate the parameters of these distributions.

The latter approaches have a direct relation to summary measures of the treatment effect and allow for a quantification of the treatment effect. 
However, there is typically no single statistic that can fully describe general differences between survival curves. The optimal choice of effect estimates may depend on the expected shape of the survival functions and on considerations which differences are clinically relevant. Conclusions regarding treatment benefit may involve different effect estimates. In particular, different stake holders, such as patients and physicians, may assign different importance to the various characteristics of the survival curve, as, e.g., improvements in median survival versus long term survival \cite{shafrin2017patient}. 
Analysis strategies in the NPH case might also be different if survival curves are crossing (e.g. more early toxic events of a beneficial treatment; frequent in paediatric oncology) than for non-crossing curves. 

The usefulness of many of the proposed alternative parameters (difference in RMST, average hazard ratio, milestone survival probabilities), as measures of the clinical benefit of an experimental treatment is controversially discussed \cite{freidlin2019methods}. For example, the restricted mean survival time, as well as the average hazard ratio, may not only be difficult to interpret, but also depend on a cut-off parameter, specifying the time span to which the mean survival time is restricted or across which the hazard ratio is averaged.  The choice of this parameter may be challenging, especially if results depend sensitively on the chosen cut-off. An important question is also if efficient, unbiased estimators and conservative confidence intervals are available, and how their properties depend on the survival distributions, patient recruitment patterns and censoring patterns.

The main objective of this paper is to assess the performance and adequacy of these methods under relevant NPH scenarios and provide recommendations on the statistical analysis and reporting of randomized clinical trials with a time-to-event endpoint under non-proportional hazards.

In preparation of this aim, we previously performed a comprehensive literature review \cite{bardo2023methods}, on methodological approaches for the analysis of time to event data under NPHs, reviewing  also the availability of statistical software that implement these methods.
This review complemented previous review articles that have focused mostly on quantitative comparisons for specific NPH scenarios, e.g. \cite{li2015statistical}, for a specific method class, e.g. \cite{rauch2018average}, or for NPH situations in specific disease areas, e.g. oncology \cite{ananthakrishnan2021critical}. 

The subsequent simulation study was planned following the clinical scenario evaluation approach by Benda et al. \cite{benda2010aspects} and Friede et al. \cite{friede2010refinement}, as well as recommendations provided in Morris et al. \cite{morris2019using}. 
It accordingly covers a broad range of underlying statistical models and parameter constellations, clinical trial design options and compares the performance characteristics of the identified statistical testing and estimation methods for the analysis of time to event data under NPH.

To ensure a neutral comparison of the identified methods, a simulation protocol was developed and made public at the European Union electronic Register of Post-Authorisation Studies (EU PAS Register, https://www.encepp.eu/encepp/studiesDatabase.jsp) before programming the simulation. The study protocol is available under the web-link https://www.encepp.eu/encepp/openAttachment/fullProtocol/49769.

The paper is structured as follows: In Section \ref{sec.methods} we introduce the notation and give a brief description of the considered analysis methods and simulation procedures. In addition, we report on the software implementations used in the simulations. In Section \ref{sec.simulation.scenarios} we specify the simulation scenarios considered (delayed onset, crossing hazards, subgroups, changing hazards after progression), describe how the effect sizes under the alternative hypotheses were chosen and based on which criteria the methods were classified. The simulation results, reporting the operating characteristics of the considered methods as well as the classification of methods, are given in Section \ref{sec.results}. In Section \ref{sec.case.study} the considered methods are illustrated in a case study. Finally, in Section \ref{sec.discussion} the findings are summarized and recommendations are discussed.

\section{Analysis methods used in the simulation study}
\label{sec.methods}

\subsection{Notation}\label{general-notation}
We consider a control and a treatment group indexed by $i=0,1$, respectively, with subscript $i$ applicable to all following notation.
Let $T$ denote a random time and $S(t)=P(T > t)$ the corresponding survival function. 
Further let $\lambda(t) = \lim_{\delta \downarrow 0} \frac{P(  T\in [t,t+\delta) | T \geq t)}{\delta}$ denote the hazard function and $\Lambda(t) = \int_0^t \lambda(s)ds$ the cumulative hazard function. 
Note that $S(t) = \exp\{-\Lambda(t)\}$. Further let $D = \{t_1,...,t_k\}$ denote the set of observed ordered unique event times and define $t_0=0$. Let $Y(t)$ be the number of subjects at risk at time $t$, $N(t)$ the cumulative number of events up to time $t$ and $dN(t)$ the number of events observed at time $t$.
We denote with $\hat{S}(t)=  \exp\{- \sum_{s\in D: s \leq t} dN(s)/Y(s)\}$ the Nelson-Aalen-Breslow estimator for the survival function \cite{Aalen.2008,aalen2010history}.
Let $\hat{S}(t)^-$ denote the left continuous version of $\hat{S}(t)$.

\subsection{(Weighted) logrank tests}
The logrank test is a widely used non-parametric test for the comparison of two (or more) survival functions. A weight function $w(t)\geq 0$ may be used to put different emphasis on different event times. 
The null hypothesis for a one-sided (weighted) logrank test comparing an experimental group to a control group is $H_0: \lambda_0(t) \leq
\lambda_1(t)$ for all $t > 0$.
The weighted logrank test statistic is
\[z = \frac{\sum_{t \in D_1 \cup D_2} w(t) (dN_0(t) - e_0(t))}{\sqrt{\sum_{t \in D_1 U D_2} w(t)^2 var(dN_0(t))}}\]
where $e_0(t)$ denotes the expected number of events in group $0$ at time $t$ and both $e_0(t)$ and the variance $var(dN_0(t))$ are calculated conditional on the numbers at risk and the number of events in both groups at time $t$ \cite{fleming1981class}. 
We consider the unweighted logrank test, i.e. $w(t) =
1$ for all $t$, and logrank tests with weights from the Fleming and Harrington (FH) rho-gamma family \cite{harrington1982class,fleming1991counting}. The rho-gamma family of weight
functions is defined as $w(t) = S^{-}(t)^{\rho} (1-S^-(t))^{\gamma}$.
In the simulation we consider parameter combinations ($\rho=1, \gamma=0$),
($\rho=1,\gamma=1$) and ($\rho=0,\gamma=1$), which respectively put more weight on early, intermediate and late event times. Note that ($\rho=0,\gamma=0$) corresponds to the unweighted logrank test.

With arbitrarily weighted logrank tests, rejection of $H_0: \lambda_0(t) \leq
\lambda_1(t)$ for all $t > 0$
does not necessarily imply that $S_1(t)>S_0(t)$ at any time-point.
(Because a lower hazard at a limited time interval does not necessarily translate to a lower cumulative hazard at any time-point.) 
Magirr and Burman \cite{magirr2019modestly} proposed a modestly weighted logrank test that 
controls the type I error rate under the null hypothesis  $H_0: S_0(t) \geq
S_1(t)$ for all $t>0$. 
For this test, the weight for the $j$-th event time is set to
\[w_{j} = \frac{1}{\text{max}\{\tilde{S}\left( t_{j - 1} \right),\tilde{S}\left( t^{*} \right)\}}\]
with $\tilde{S}(t)$ denoting the Kaplan-Meier estimate at time \(t\) based on
the pooled data from both treatment arms. Observations before time $t^*$ all get the same (low) weight, and starting from $t^*$, weights are increasing. In the simulation, $t^*=6$ months and $t^*=12$ months was used.

Note that the unweighted logrank test is a special case of the modestly weighted test and thus allows for the same extended conclusion regarding the survival functions.

The MaxCombo test is a combination of a set of weighted logrank tests \cite{tarone1981distribution,lee2007versatility,fleming2011counting,karrison2016versatile,ristl2021delayed}. In brief, a set of different weight functions is defined and a weighted logrank test is performed for each weight function. The test statistics for MaxCombo test is the maximum across the test statistics of these weighted tests. The p-value is calculated based on a multivariate normal approximation for the joint distribution of these statistics, i.e. a parametric maximum type multiplicity adjustment across the included tests is performed. The MaxCombo test is intended to be used when the best weighting scheme is not known a priori. In the simulation, we considered the combination of tests with rho-gamma family weight parameters ($\rho=0, \gamma=0$), ($\rho=1,\gamma=0$), ($\rho=1, \gamma=1$), ($\rho=0, \gamma=1$).

\subsection{Restricted mean survival times}
The restricted mean survival time (RMST) corresponds to the area under the survival curve up to a predefined cut-off time-point \cite{royston2011use,hasegawa2020restricted}. A difference in restricted mean survival times between experimental treatment and control can be interpreted as the average survival time gained in the time interval up to the cut-point.

The RMST in group $i$ up to a pre-specified time-point $L$ is $\mu_i=\int_0^L S_i(t)dt$. The corresponding non-parametric estimate used in the simulation is the corresponding area under the estimated survival curve. Formally, let $D'_i$ be the number of unique event times $t_{i,1} < \hdots < t_{i,D'_i} \leq L$ 
in group $i$ that are less or equal $L$. Further define $t_{i,0}=0$ and $t_{i,D'_i+1}=L$ and $\Delta t_{i,j}= t_{i,j+1}-t_{i,j}$. The according estimate for the RMST in group $i$ is $\hat\mu_i=\sum_{j=0}^{D'_i} \hat S_i(t_j)\Delta t_{i,j}$ and the estimated RMST difference between the two groups is $\hat\mu_1-\hat\mu_0$. $\hat\mu_i$ is asymptotically normally distributed around $\mu_i$ and an estimate for the variance is given by
\[\hat{var}(\hat{\mu_i})= \sum_{u=1}^{D'_i} \left(\sum_{j=u}^{D'_i} \hat S_i(t_j)\Delta t_{i,j}\right)^2 \frac{dN_i(t_u)}{Y_i^2(t_u)}\]
The null hypothesis $H_0: \mu_1 - \mu_0 \leq 0$ was tested by a Wald test based on the normal approximation of $\hat{\mu_0}$ and $\hat{\mu_1}$. Similarly, Wald confidence intervals for the difference $\mu_1 - \mu_0$ were calculated. 

In the simulation, RMSTs were calculated for the first 6 and 12 months, i.e. $L=6$ and $L=12$.

\subsection{Median survival time}\label{difference-in-median-survival-time}
Median survival time in group $i$ is defined as $\tau_i = inf(t: {S_i}(t) \leq 0.5)$

The median was estimated non-parametrically from the empirical survival function as $\hat{\tau_i} = inf(t: \hat{S_i}(t) \leq 0.5)$. The estimate is asymptotically normally distributed around ${\tau}_{i}$. A variance estimator for $\hat{\tau_i}$ is derived from the counting process representation of event times \cite{fleming1991counting,ristl2023simultaneous} as $\left(\frac{1}{-\hat\lambda(\hat{\tau_i})}\right) ^2 \sum_{s \in D_i: s\leq \hat{\tau_i}} \frac{dN_i(s)}{Y_i^2(s)}$. Here, $\hat\lambda(\hat{\tau})$ is an estimate for the hazard rate at $\hat{\tau}$, which may be obtained by Kernel density estimation. 

As an alternative method, parametric Weibull models were fit to each group separately and an estimate for the median was obtained as plug-in estimate from the estimated scale and shape model parameters. The variance of the estimated median was calculated from the estimated covariance matrix for the scale and shape estimates using the delta method.

The null hypothesis $H_0:\tau_1-\tau_0 \leq 0$ was tested by a Wald test based on the normal approximation of $\hat{\tau}_i$.  Similarly, Wald confidence intervals for the difference $\tau_1-\tau_0$ were calculated.

\subsection{Milestone survival probabilities}\label{milestone-survival-probabilities-based-on-km}
The null hypothesis for a difference in milestone survival probabilities $H_0: S_1(t) \leq S_0(t)$, for a predefined time $t=6$ months and $t=12$ months, was tested by Wald tests based on the normal approximation for $\hat{S}_i$ \cite{fleming2011counting,ristl2023simultaneous}. Similarly, Wald confidence intervals were calculated.

\subsection{Accelerated failure time model}
In an accelerated failure time (AFT) model for the comparison of the two groups, it is assumed that the survival times $T_1$ have the same distribution as $T_0*\exp(\theta)$, where $\exp(\theta)$ is the acceleration factor \cite{klein2003survival}. Equivalently, $\log(T_1)$ has the same distribution as $\log(T_0)+\theta$ and the resulting regression model is $\log(T) = \mu+\theta*x+\sigma W$, where $\mu$ is an intercept term, $x=0,1$ indicates the treatment group, $\sigma$ is a scale factor and $W$ is a random error term for which a certain distribution is assumed. In the simulation we explored the lognormal and the Weibull AFT model, i.e. $W$ was assumed to follow either a $N(0,1)$ distribution or an extreme value distribution with density $f(x)=\exp(x-\exp(x))$.
Model parameters were estimated using the maximum likelihood method and model based standard errors were calculated.
The null hypothesis $H_0: \theta \geq 0$ was tested as Wald test for the respective model coefficient.

\subsection{Average hazard ratio}
An average hazard ratio can be defined as 
\[\theta = \frac{\int_0^L W(s) d{\Lambda}_1(s) }{\int_0^L W(s) d{\Lambda}_0(s) }\]
where $L$ is a predefined cut-off time point and $W(s),s\geq 0$ is a non-negative monotonically decreasing weight function with values in $[0,1]$ \cite{kalbfleisch1981estimation,rauch2018average}. In the simulation study, we consider the average hazard ratio with weight function $W(s)= {S}_0(s){S}_1(s)$ and its corresponding estimate $\hat{W}(s)=\hat{S}^{-}_0(s)\hat{S}^{-}_1(s)$. Here the left continuous estimator of the survival function, $\hat{S}^{-}$, is used to obtain a predictable function, which is a formal requirement to establish asymptotic arguments in the counting process framework.
The average hazard ratio based on the considered weight function is identical to $\frac{P(T_1 \wedge L >T_0 \wedge L)}{P(T_1 \wedge L< T_0 \wedge L)}$ and can be interpreted as a type of concordance statistic. Unlike the Cox model hazard ratio estimate, the limiting value of this average hazard ratio estimate does not depend on the censoring distribution. Under proportional hazards, the average hazard ratio and the Cox model hazard ratio coincide.

Inference is based on the log-average hazard ratio, which is estimated as 
\[{\widehat{\log \theta}} =\log\left(\int_0^L \hat{W}(s) d\hat{\Lambda}_1(s) \right)-\log\left(\int_0^L \hat{W}(s) d\hat{\Lambda}_0(s) \right)\]
An estimate for the variance of the contribution of group $i$, $\log\left(\int_0^L \hat{W}(s) d\hat{\Lambda}_i(s) \right)$, results from the counting process representation \cite{fleming1991counting,ristl2023simultaneous} as
\[
\hat\nu_i = \left( \frac{1}{\log\int_0^L\hat{W}(s)d\hat\Lambda_i(s)} \right) ^2 \sum_{s \in D_i: s\leq L}\hat{W}^2(s) \frac{dN_i(s)}{Y_i^2(s)}
\]
and the variance of $\widehat{\log \theta}$ can be estimated by $\hat\nu_0+\hat\nu_1$.

The log average hazard ratio estimate is asymptotically normally distributed around the true log average hazard ratio. Accordingly, the null hypothesis $H_0: \theta \geq 1$ or equivalently $H_0: \log \theta \geq 0$ was tested by a Wald test based on the normal approximation of $\widehat{\log \theta}$. Similarly, Wald confidence intervals for the log average hazard ratio were calculated. Wald confidence intervals for the average hazard ratio were obtained by transforming the confidence limits of the log average hazard ratio.

In the simulation, the average hazard ratio was calculated for the first 6 and 12 months, i.e. $L=6$ and $L=12$.

\subsection{Software implementation used in the simulation}
In the simulation study, the logrank test was computed using the function
survdiff in the R library survival. Rho-gamma weighted logrank tests were
calculated using the function logrank.test in the R library nph. The modestly weighted logrank test was calculated using the function wlrt in the R library nphRCT. The function logrank.maxtest from the R library nph was used in the
simulation study to calculate the MaxCombo test.
The function nphparams in the R library nph was used for estimation and statistical inference of differences in RMSTs, median survival times, milestone survival probabilities and average hazard ratios.
The function survreg of the R package survival was used to fit accelerated failure time models and Weibull hazard models.

\section{Simulation scenarios}\label{sec.simulation.scenarios}
To assess the operating characteristics of the analysis methods described in Section \ref{sec.methods} under different settings with non-proportional hazards, we performed a simulation study with four classes of scenarios comprising (i) delayed onset of treatment effect, (ii) crossing hazard curves, (iii) presence of a subgroup defined by a baseline biomarker where the treatment effect is different, and (iv) a change in hazards after disease progression. In each class of scenarios we varied relevant design characteristics such as the time-point of changing hazards, the proportion of sub-groups or recruitment rates.
Random censoring was assumed to follow an independent exponential distribution.

The full simulation study included 10608 different scenarios (1296 with delayed onset, 1824 with crossing hazards, 5760 with biomarker subgroups and 1728 with disease progression), by combining different settings for sample size, recruitment rate, targeted power, piece-wise constant  hazard rates under control, piece-wise constant hazard rates, censoring rates, progression rates, and subgroup proportions. To allow for a concise presentation, we focus on a subset of representative parameter settings in the paper while results on all investigated scenarios are available in the full simulation report \cite{confirms2023report}, which is available as online supplementary material.

In the paper, we assumed  for all scenarios a sample size of up to 500 patients across groups who are recruited with constant recruitment rate over 18 months, a target number of events of 375 (corresponding to 75\% of the maximal number of included subjects), a target power of 80\% (see Section \ref{sec.calibration} for details) and 10\% random censoring (see Section \ref{sec.censoring}).

For each scenario, 2,500 simulation runs were performed and the analysis methods described in Section \ref{sec.methods} were applied. The empirical power, confidence interval coverage and bias of parameter estimates were calculated from the simulation results.

Scenarios in which both groups were sampled under assumptions of the control were included, too, to assess type I error rates under the null hypothesis of equal survival functions. 

\subsection{Delayed onset}
In scenarios with delayed onset, a constant hazard rate of 0.693 per year, corresponding to a median survival time of 12 months, was assumed in the control group. In the treatment group, the same hazard rate was assumed until a time-point of 0, 2, 4, 6 or 8 months. The respective hazard ratios after this delay time were 0.75, 0.71, 0.67, 0.60 and 0.50.
The resulting survival, hazard and hazard ratio functions are shown in Supplementary Figure \ref{fig.scenario_delay}.

\subsection{Crossing hazards}
Crossing hazard functins, i.e. $\lambda_1(s) > \lambda_0(s)$ and $\lambda_1(s') < \lambda_0(s')$ for some time points $s \neq s'$, may occur if a control treatment is superior to experimental treatment early after treatment initiation while the experimental treatment is superior in the long run. 
A constant hazard rate of 0.693 per year was assumed for the control group,  
and piece-wise constant hazards were assumed for the treatment group, with a change in hazards after 0, 2, 4, 6 or 8 months. Hazards were larger under treatment than under control before this time-point, with a hazard ratio of 1.5, and lower afterwards with hazard ratios 0.75, 0.64, 0.50 and 0.30, respectively. Additionally, a scenario with a hazard ratio of 3 in the first two months and a hazard ratio of 0.43 afterwards was included. 
The resulting survival, hazard and hazard ratio functions are shown in Supplementary Figure \ref{fig.scenario_crossing}.

\subsection{Biomarker subgroups}
In the biomarker subgroup scenarios, we assumed the population is comprised of two subgroups, biomarker positive and biomarker negative subjects, who have the same survival distribution under control but may respond differently to the treatment. In this setting, in the overall population non-proportional hazards occur as the composition of the overall population changes with time.
For both subgroups, we assumed a constant hazard rate of 0.693 per year under control. Under treatment, the assumed hazard ratios between biomarker positive and biomarker negative patients were 0.3, 0.7, 0.8 and 0.9, corresponding to a stronger treatment effect for biomarker positive patients. The prevalence of biomarker positive patients took values of 0.1, 0.3, 0.5, 0.7 or 0.9. Hazard rates for the biomarker negative subgroup were calculated depending on the combination of all other parameters according to Section \ref{sec.calibration}.
The resulting survival, hazard and hazard ratio functions are shown in Supplementary Figure \ref{fig.scenario_subgroups}.

\subsection{Changing hazards after progression}
In the scenarios with disease progression, we assume that hazards and the relative treatment effect change after progression. Here, survival is modelled in terms of a multistate model with a pre-progression state, a post-progression state and death as absorbing state. Possible transitions are from pre-progression directly to death or to post-progression and subsequently from post-progression to death. Transitions between states were modelled with constant hazard functions, with hazard values depending on the current state and the treatment group. 

In the simulation, this model was parameterized by the median survival time set to 12 months under control, the proportion of subjects expected to progress before death in either treatment group with values of 0.1 or 0.2. Following, progression subjects reverted to the same hazard in both groups, which was either 1.25 or 2 times the pre-progression hazard in the control group, i.e. the hazard was increasing after progression in both groups, though possibly to a different extent. We only considered scenarios where the progresion rate in the control group was as large or larger than that of the treatment group (i.e., 0.1 and 0.1, 0.1 and 0.2, 0.2 and 0.2 in the treatment and control group). For scenarios with a 0.1 and 0.2 progression rate in the treatment and the control group one may imagine a treatment that acts on progression but not necessarily (pre-progression) survival. In the paper we consider scenarios with equal median survival between group, as well as, scenarios with a target power of 80\%.  
The absolute hazard values for the treatment group were calculated according to the calibration described in Subsection \ref{sec.calibration}.
In contrast to the previous scenarios, the disease progression scenarios include settings where the hazard ratios move closer to 1 with increasing event time.
The resulting survival, hazard and hazard ratio functions are shown in Supplementary Figure \ref{fig.scenario_progression}.

\subsection{Calibration of the assumed treatment effects}\label{sec.calibration}
To focus the assessment of the different methods on a power range around 80\%, the between-group differences in the simulation scenarios were calibrated. For this purpose, the parameters were chosen to calibrate the difference in median survival between groups, such that under proportional hazards the power would be 80\% for a logrank test at the one-sided 2.5\% significance level.

In detail, given the sample size, recruitment rate, target number of events, the piece-wise constant hazard rates under control, censoring rates, subgroup proportions, and, for scenarios with two time-intervals with different hazards, the hazard ratio in the first time interval, the median survival time in the control group was determined. Then, the piece-wise constant hazard rate in the second time-interval of the treatment group group or, in case of constant hazards, the hazard rate under treatment was chosen such that the two medians when applied under the assumption of corresponding constant hazards to Schoenfeld's sample size formula \cite{schoenfeld1981asymptotic} gave 80\% power at the one-sided 2.5\% significance level.

\iffalse
\begin{table}[ht]
\centering
\caption{Simulation standard error and corresponding margins derived as multiples of the corresponding simulation standard error for the different effect size scenarios, defined with respect to the power of the log-rank test with corresponding difference in median survival times under proportional hazards.}
\label{tab:margins}
\begin{tabular}{|c|c|c|c|}
\hline
Rejection Rate & SSE & Factor & Margin \\
\hline
0.0250 & 0.0031 & 2.3000 & 0.0072 \\
0.5000 & 0.0100 & 4.0000 & 0.0400 \\
0.8000 & 0.0080 & 4.0000 & 0.0320 \\
0.9000 & 0.0060 & 4.0000 & 0.0240 \\
\hline
\end{tabular}
\end{table}

\fi

\subsection{Random Censoring}\label{sec.censoring}

Random censoring was simulated from an exponential distribution with some parameter $\lambda_c$. To define  $\lambda_c$, we first define a cutoff \(t_{max}\) set as the 1/10000 quantile of the survival function of
the treatment or control arm (whichever is larger).  Now, $\lambda_c$ is chosen such that, before \(t_{max}\), in the absence of administrative censoring, the expected proportion of patients whose survival time is censored before the event occurs is  $p$. 

Since
\begin{equation}
p = \frac{1}{2} \cdot\frac{\Lambda_c(t_{max})}{\Lambda_c(t_{max}) + \Lambda_{0,evt}(t_{max})} + \frac{1}{2}\cdot\frac{\Lambda_c(t_{max})}{\Lambda_c(t_{max}) + \Lambda_{1,evt}(t_{max})}\label{eq:cens}
\end{equation}

\noindent where the cumulative hazard for random censoring is denoted by \(\Lambda_c\) and the cumulative
hazard for the event of interest is denote by \(\Lambda_{0,evt}\) and \(\Lambda_{1,evt}\) in the
control and treatment arm, respectively.
Now, $\lambda_c$ is obtained by  setting 
$\Lambda_c(t_{max})=\lambda_c \cdot t_{max}$ 
in (\ref{eq:cens}) and solving (\ref{eq:cens}) for $\lambda_c $.

\subsection{Summary of results}
To summarize the results for each considered parameter constellation we graded the hypothesis testing and the estimation procedures according to their T1E control, power, and confidence interval coverage. Then we reported for each method the frequency distribution of the grades across the considered parameter constellations. While this gives a coarse summary of the methods' performance the specific distribution depends on the specific choice of simulation scenarios and the grading scheme.  
\paragraph{Classification of hypothesis tests.} To compare the power of different hypothesis tests across all considered scenarios in the full simulation study, we defined the  following grading schemes.

To assess power relative to the logrank test, we graded methods “*” if they provided a power advantage in excess of four simulation standard errors when compared to the log-rank test.  Methods were graded “+” if their power was within plus minus four simulation standard errors of the log-rank test’s power, “$\sim$” if power was more than four simulation standard errors below the power of the log-rank test, and “-” if power was more than eight simulation standard errors below that of the log-rank test.
    
To assess power relative to the best performing method, methods were graded “+” if their power was equal or no less than four simulation standard errors compared to the best performing method for a given scenario. We assigned “$\sim$” if a method provided marginally less power than the best performing method (i.e. between four and eight simulation standard errors less power) and “-” if a method provided substantially less power than the best performing method (i.e. more then eight simulation standard errors). In addition we assigned “*” to the method with the best performing method in each scenario. Note, that category “top” (*) does not imply that the respective method outperformed the remaining methods by any given margin, but just that it numerically provided the most power in a given setting. The next best performing methods may still provide comparable powers with respect to the applied margins. In case that powers were identical between methods (e.g. 100\% for certain scenarios) all corresponding methods were graded “top”.

\paragraph{Classification of estimates.}
Concerning estimation, we graded methods according to bias and coverage probability of confidence intervals across all scenarios in the full simulation study. 
Methods were graded “*” if the coverage probability was within plus/minus 2.3 simulation standard errors of the nominal coverage. Methods with coverage probabilities more than 2.3 simulation standard errors below the nominal level were graded “-”, methods with coverage more than 2.3 simulation standard errors above the nominal level were graded “+” indicating conservative control of the coverage probability. The factor 2.3 was chosen because it corresponds approximately to the 99\% quantile of the standard normal distribution. With 2,500 simulations and a nominal coverage of 95\%, 2.3 times the simulation standard error corresponds to a deviation of one percentage point.

For bias, deriving acceptance margins in terms of the simulation standard error did not appear practical. For example, for the difference in median survival in the delayed effect scenarios the estimated simulation standard error was between two and five days for actual differences of up to around 600 days (about half of the control group median survival of 36 months). A corresponding margin of 10 days would probably not represent a relevant difference in performance. Furthermore, the standard error depends on the sampling distribution of the estimator, the scale of the estimand, and varies strongly across different parameter settings beside the effect size scenario (e.g. sample size). Consequently, we use the following margins:

For the average hazard ratio, estimated biases within plus minus 0.05 were considered good (+), positive biases (i.e. conservative with respect to the one-sided null hypothesis) above 0.05 were considered average ($\sim$) and negative biases above 0.05 (i.e anti-conservative with respect to the one-sided null hypothesis) were considered poor (-). Consequently, in a scenario with true average hazard ratio of 1 we would consider estimates between 0.95 and 1.05 acceptable, estimates above 1.05 as average, and estimates below 0.95 as poor.
    
For methods estimating differences in median survival or RMST, bias was assessed relative to the corresponding control group parameter value. E.g. for scenarios with a control group median survival of 12 months, an absolute bias in the estimate of the difference of median survival between groups of 3 months corresponds to a relative bias of 3/12, or 25\%. With that, we classified the estimated bias of methods as good (+) if it was between plus/minus five percentage points, and average ($\sim$) if the estimated bias was negative (i.e. conservative with respect to the one-sided null hypothesis) or poor (-) if the bias was positive.

While these margins may be arbitrary, they are larger than several simulation standard errors across all scenarios and do permit an interpretation in terms of differences that would be considered unlikely to arise due to the sampling variation of the simulation study. At the same time, the chosen margins in combination with relatively small simulation standard errors are not prohibitively large such that a differentiation between methods with different operating characteristics is possible.

\section{Results\label{sec.results}}
We present results for the selected scenarios as described in Section \ref{sec.simulation.scenarios}. A summary of results from the full simulation study, which includes additional scenarios can be found in the full simulation report \cite{confirms2023report}, which is included in the online supplementary material.

\subsection{Type I error rate}
In the scenarios with delayed onset, crossing hazards and biomarker subgroups, the survival function in the control  group (and, under the null hypotheses, also in the treatment group) has a constant hazard rate corresponding to a median survival of 12 months. Therefore we report the pooled simulation results from these scenarios with a corresponding larger number of simulation runs in the assessment of the type I error rate. In the disease progression scenario, the control hazard function is monotonically increasing with time, as the hazard rate increased after progression. Here, a scenario with median survival time of 12 months was chosen, corresponding to a proportion of subjects expected to progress of 0.2 and a hazard ratio of 0.5 between post-progression and pre-progression. Type I error rates were simulated under equal survival functions in both treatment groups, corresponding to these two settings (see  Table \ref{tab.T1E}).

For all methods, the type I error rate was close to the nominal level of 2.5\%. This result is in line with the larger set of simulations (see the full simulation report \cite{confirms2023report})
where in general control of type I error rate was observed for all methods.

\begin{table}[htbp]
  \centering
  \caption{Empirical type I error rates (T1E) from 12,500 simulation runs under equal survival functions in a scenario with constant hazards and 2,500 simulation runs for a scenario with increased hazards after disease progression.}
    \begin{tabular}{lrr}
    \hline
    Method & T1E constant hazard scenario & T1E progression scenario\\
    \hline
    AHR 6m & 2.45  & 2.32 \\
    AHR 12m & 2.63  & 2.60 \\
    Milestone surv. 6 months & 2.46  & 2.24 \\
    Milestone surv. 12 months & 2.53   & 2.32 \\
    RMST difference 6 months & 2.30  & 2.44 \\
    RMST difference 12 months & 2.42  & 2.24 \\
    AFT Weibull & 2.63  & 2.56 \\
    AFT lognormal & 2.70   & 2.20 \\
    Median difference & 2.37 & 2.60 \\
    Weibull median difference & 2.39  & 2.48 \\
    FH 0-1 & 2.66  & 2.44 \\
    FH 1-0 & 2.46  & 2.68 \\
    FH 1-1 & 2.69  & 2.76 \\
    Logrank & 2.55  & 2.64 \\
    MaxCombo & 2.61  & 2.48 \\
    Modestly weighted 6 months & 2.54  & 2.72 \\
    Modestly weighted 8 months & 2.58  & 2.80 \\
    \hline
    \end{tabular}
  \label{tab.T1E}
\end{table}

\subsection{Power of hypothesis tests}

We evaluate the power of Fleming-Harrington weighted logrank tests, the MaxCombo test and modestly weighted logrank tests, as well as the power of hypothesis tests based on estimates for the average hazard ratio, differences in milestone survival probabilities, RMST and median survival as well as the acceleration factors from Weibull and lognormal AFT models. 

\subsection{Delayed onset}
\begin{figure}[!htbp]
\begin{center}
\includegraphics[scale=.55,clip,trim=0cm 0cm 0.00cm 0cm]{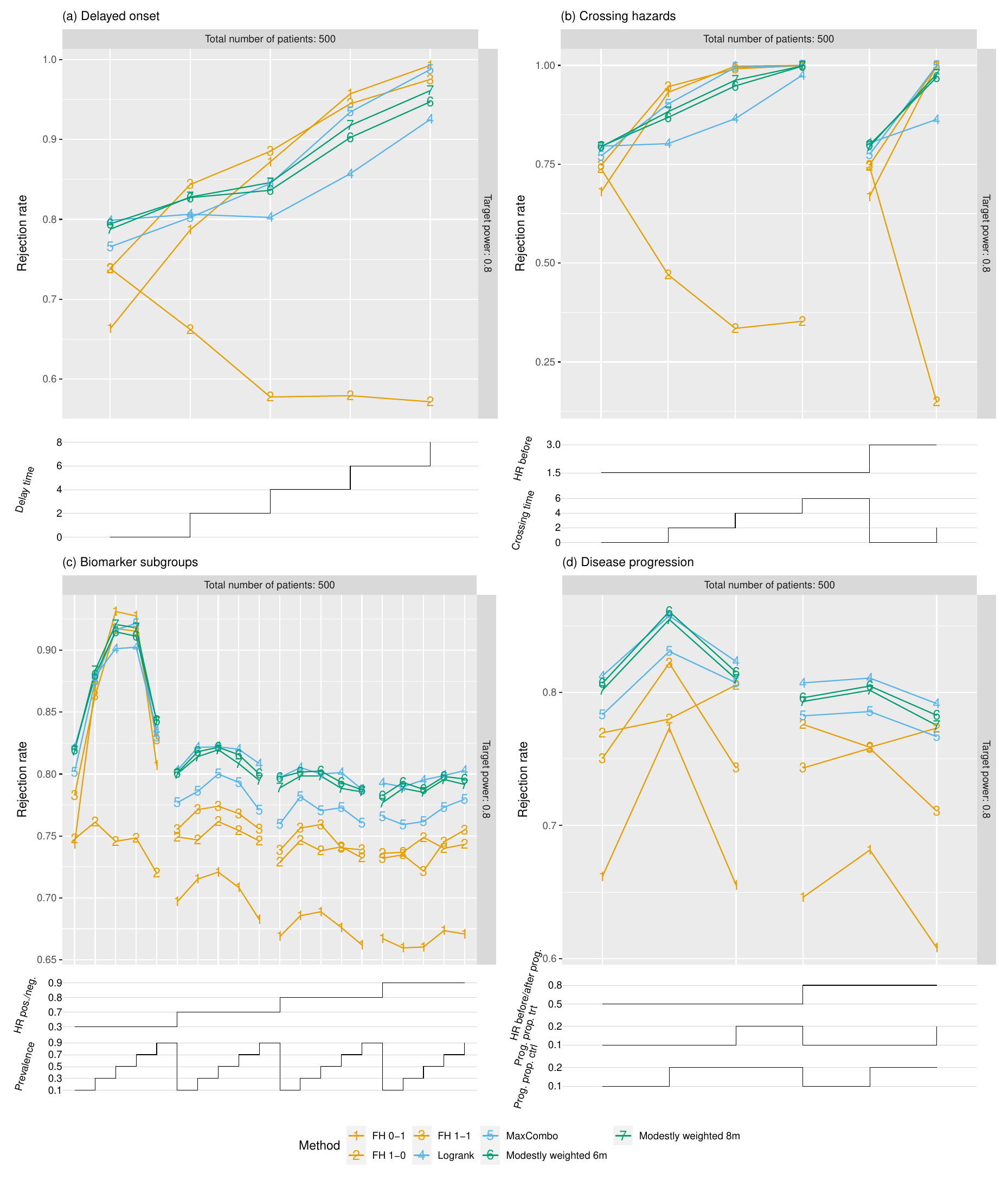}
\end{center}
\caption{Power of studied non-parametric hypothesis tests in the scenarios with (a) delayed onset of treatment effect, (b) crossing hazard, (c) biomarker subgroups and (d) disease progression. For all scenarios, rejection rates are plotted versus specific scenario settings. The chosen parameter values, which vary within a scenario class, are displayed as step functions underneath the respective main plot. The step functions are understood to be right-continous, i.e. reading the figure from left to right, at each step the new value counts.
}
\label{fig.power_tests}
\end{figure}

Figure \ref{fig.power_tests}a shows the power of the different test procedures under scenarios with delayed onset. Note that due to the calibration (see Subsection \ref{sec.calibration}) of treatment effects, the difference in median survival is constant across scenarios and therefore a larger delay time results in a  disproportionately stronger treatment effect after the delay. Therefore, depending on the test procedure, a larger delay may not necessarily result in a lower power.
In contrast, except for the Fleming-Harrington test with weights (1,0), which emphasises early events, this stronger effect at later time-points results in increasing power. Also with increasing delay, the power advantage of weighted tests that prioritize later differences increases compared to the logrank test .
 The Fleming-Harrington test with weights (0,1) provides the largest power for settings with long delays. However, it is also the worst performing method when there is no delay in treatment effect onset, and improves on the (unweighted) logrank test only with delays of 4 months and more.
  The Fleming-Harrington test with weights (1,1) provides good power across scenarios with moderate to long delays, however it incurs a substantial disadvantage in settings without a delay.
Similarly, the MaxCombo test provides high power across a broad range of settings, though with moderate losses in settings without delay.
The modestly weighted procedures provide moderately improved power compared to the logrank test in settings with delayed treatment effect onset, while incurring only limited losses in settings without delay.

\begin{figure}[!htbp]
\begin{center}
\includegraphics[scale=.55,clip,trim=0cm 0cm 0.00cm 0cm]{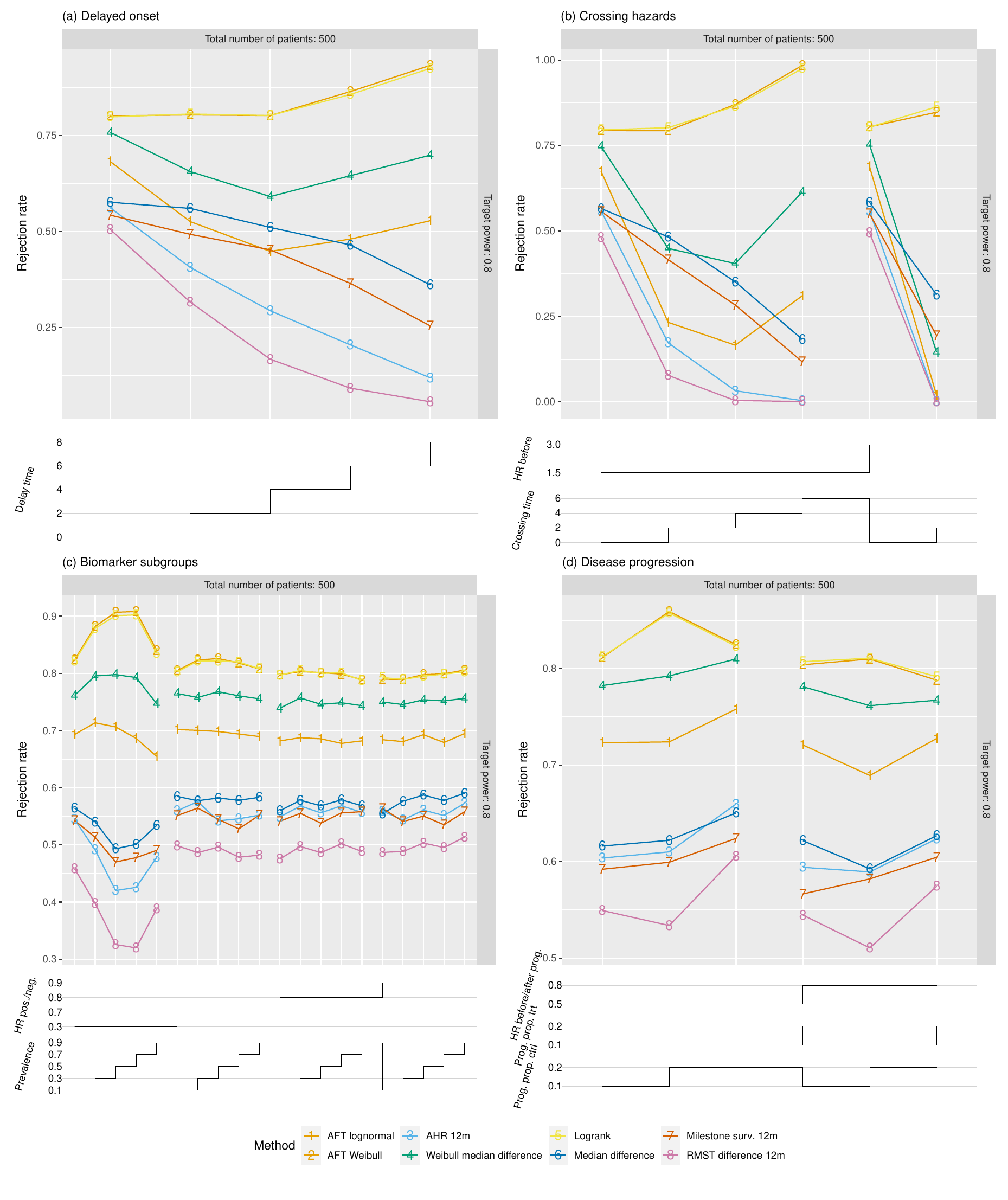}

\end{center}
\caption{Power of hypothesis tests derived from different parameter estimates in the scenarios with (a) delayed onset of treatment effect, (b) crossing hazard, (c) biomarker subgroups and (d) disease progression. For all scenarios, rejection rates are plotted versus specific scenario settings. The chosen parameter values, which vary within a scenario class, are displayed as step functions underneath the respective main plot. The step functions are understood to be right-continuous, i.e. reading the figure from left to right, at each step the new value counts.}
\label{fig.power_estimates}
\end{figure}

\begin{figure}[!htbp]
\begin{center}
\includegraphics[scale=.55,clip,trim=0cm 0cm 0.00cm 0cm]{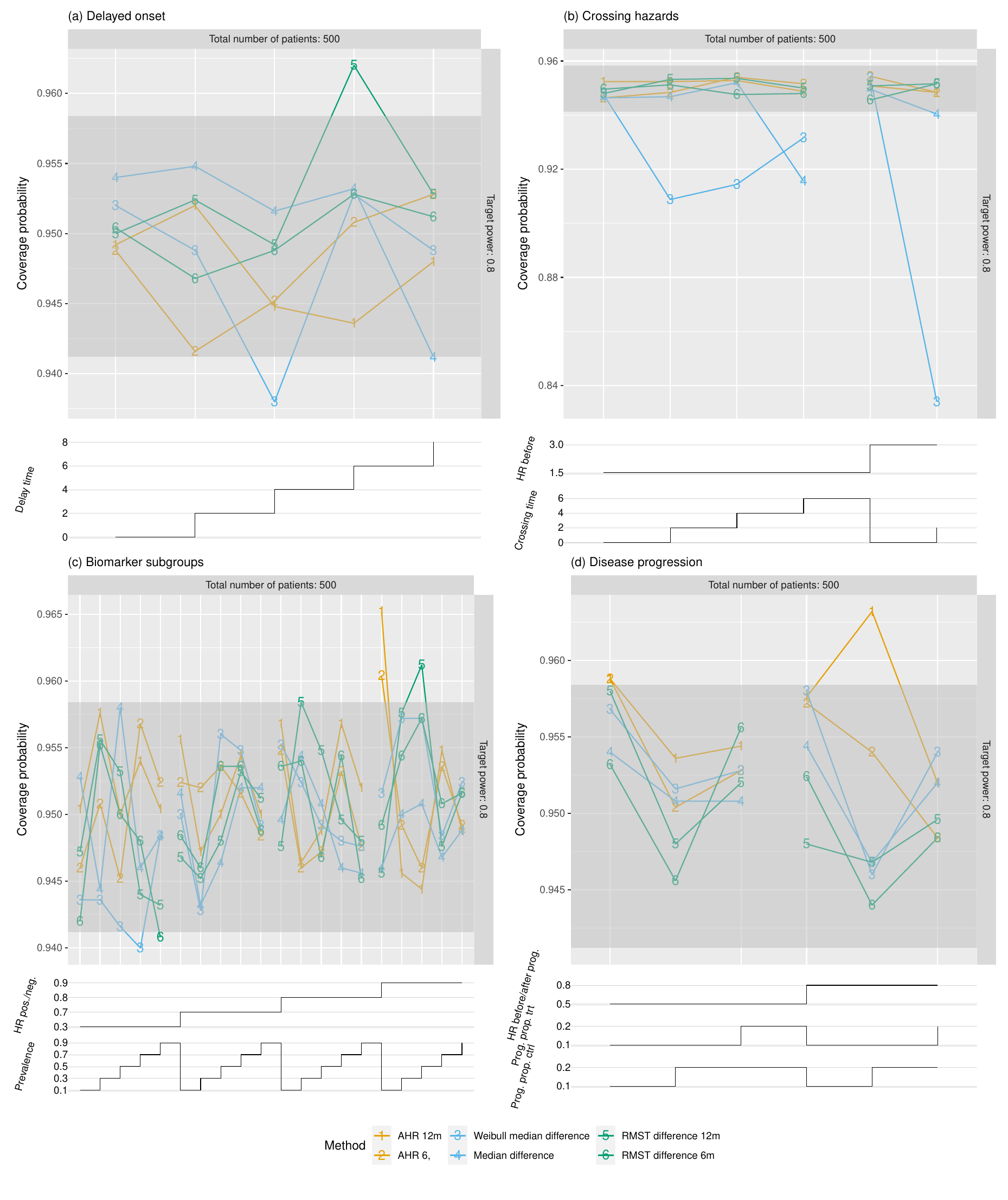}

\end{center}
\caption{Confidence interval coverage for different parameters scenarios with (a) delayed onset of treatment effect, (b) crossing hazard, (c) biomarker subgroups and (d) disease progression. For all scenarios, the empirical coverage probabilities are plotted versus specific scenario settings. The chosen parameter values, which vary within a scenario class, are displayed as step functions underneath the respective main plot. The step functions are understood to be right-continuous, i.e. reading the figure from left to right, at each step the new value counts.
The dark grey regions indicate the nominal coverage probability of 95\% $\pm$ 1.96 times the simulation standard error for an interval with exactly 95\% coverage.
}
\label{fig.coverage}
\end{figure}

Figure \ref{fig.power_estimates}a shows the power of hypothesis tests based on parameter estimates. The logrank test is included for comparison.
The logrank test and the AFT Weibull model demonstrate nearly identical power, while the difference in median survival based on the Weibull model follows closely but experiences a significant power loss. All other methods exhibit substantially lower power. This outcome can be explained by the fact that most of the considered parameters, such as the 12-month RMST,
are more sensible to detect early differences in survival curves, which results in low power in a setting of delayed onset of treatment effect.
The AFT lognormal model performs worse than the Weibull model. This may be due to deviations from the model's distributional assumptions and  may indicate a lack of robustness of this model.
Confidence interval coverage was close to the nominal level, taking into account simulation error,  for all methods, see Figure \ref{fig.coverage}a.

\clearpage

\clearpage

\clearpage

\clearpage

\subsection{Crossing hazards}

In the crossing hazards scenarios, the power of the studied hypothesis tests in general increases with increasing crossing time due to the calibration of the treatment effect (see Subsection \ref{sec.calibration}), resulting in larger survival advantages following the change in treatment arm hazard required to achieve comparable median survival differences, see Figure \ref{fig.power_tests}b.
As expected, the Fleming-Harrington test with weights (1,0) has in general low power, because the treatment effect points in the wrong direction at early time-points.
In this setting, also the unweighted logrank test has considerable less power than the other weighted logrank tests. And among these, the Fleming-Harrington tests with weights (1,1) provides slightly more power than the Fleming-Harrington test with weights (0,1). The modestly weighted tests have power between the Fleming-Harrington (0,1) test and the unweighted logrank test.
The logrank test and the AFT Weibull models provide very similar power, which is close to the
target power under proportional hazards and for moderate positive crossing times. For these procedures, the power increases with increasing crossing times (and the entailed stronger effect after crossing).
For all other tests based on parameter estimates, the power is substantially lower than the target power in settings with crossing hazards see Figure \ref{fig.power_estimates}b.

Tests targeting estimates over a restricted time-span, such as 12-months RMST or 12 months survival differences, depend on the fraction of time-points before (with detrimental effect) and after crossing (with beneficial effect) and accordingly approach low power for later crossing times. Tests for median survival differences based on the Weibull model show larger power than tests using the median derived from the empirical survival
function. A possible bias of the Weibull model, however, will be discussed below.
Confidence intervals had the desired coverage up to simulation error across methods, except for the confidence interval for the  difference in median survival, see Figure \ref{fig.coverage}b. The confidence intervals for the median difference based on the Weibull model showed severe undercoverage in all scenarios with crossing times greater than 0, with empirical coverage below 84\% in one instance.  Closer inspection of the simulation results suggests that this is due to bias of the corresponding estimator, which possibly results from the mismatch between the true distribution and the model assumption of a Weibull distribution. In all scenarios, the true median difference was 123 days, and the empirical relative bias of the Weibull-based median estimate was between -32\% and 44\% in scenarios with crossing hazards. The non-parametric estimate of the median had a small empirical bias of approximately 5\% in the two scenarios with lower-than nominal coverage. However, for the non-parametric method, the bias was reduced in scenarios with larger sample size that were assessed in the full simulation study, and coverage was within simulation error around the nominal value. In contrast, bias and undercoverage were aggravated for the Weibull-based median with larger sample size.

\subsection{Biomarker subgroups}
In the subgroup scenarios, the power of tests putting more weight on later event times is larger than for the log-rank test if the subgroup effect is very strong. In settings where the subgroup effect is moderate, and as a result the deviation from the proportional hazards assumption is minor, the logrank test has the highest power, followed by the modestly weighted tests and the max-combo test, see Figure \ref{fig.power_tests}c.

As in the other scenarios, the logrank test and the AFT Weibull model
have the largest power compared to other tests based on parameter estimates. The tests for the difference in median survival, milestone survival and RMST have
substantially lower power, see Figure \ref{fig.power_estimates}c.
Confidence interval coverage was close to the nominal value across methods, see Figure \ref{fig.coverage}c.

\subsection{Disease progression}
Overall, the logrank test appeared to provide the largest power in the disease progression scenarios, the Fleming-Harrington test with weights (0,1) had the lowest power, whereas the Fleming-Harrington test with weights (1,0) provided a power close to the logrank test, see Figure \ref{fig.power_tests}d. As in other scenarios, the MaxCombo test was largely unaffected by including one none-efficient test and maintained power close to the logrank test. 
  Notably, the modestly weighted test provide similar power as the logrank test. This complements results from other scenarios, where we have seen that the modestly weighted tests incur minimal losses compared to the logrank test when hazards are close to proportional.

As for the other scenarios, the logrank test and the AFT Weibull model performed similar and had substantially larger power than the tests based on parameter estimates. Among these, the difference in 12-month RMST was the least powerful and the tests for the median difference, 12-month milestone survival and average hazard ratio performed similar with slightly larger power. See Figure \ref{fig.power_estimates}d.
As in the other scenarios, confidence interval coverage was close to the nominal level within a range expected due to simulation error (Figure \ref{fig.coverage}d).

\subsection{Performance grading of methods}\label{sec.grading}
The results from the simulation scenarios presented above are in line with the larger set of results from the full simulation study reported in \cite{confirms2023report}. In this section we summarize the results including the scenarios considered only in the full simulation study.  Aspects only covered there were different amounts of censoring (ranging from 0 to 30\%), different baseline hazards 
(corresponding to median survival times of 6 and 36 months), additional target powers (50\% and 90\%), and one additional recruitment rate (36 months). 
However, these had little impact on the ranking of methods with respect to power, type I error rate, confidence interval coverage or bias. The amount of deviation from the proportional hazards assumption, in contrast, impacted the power differences between methods.

\begin{figure}[ht]
\includegraphics[scale=.55,clip]{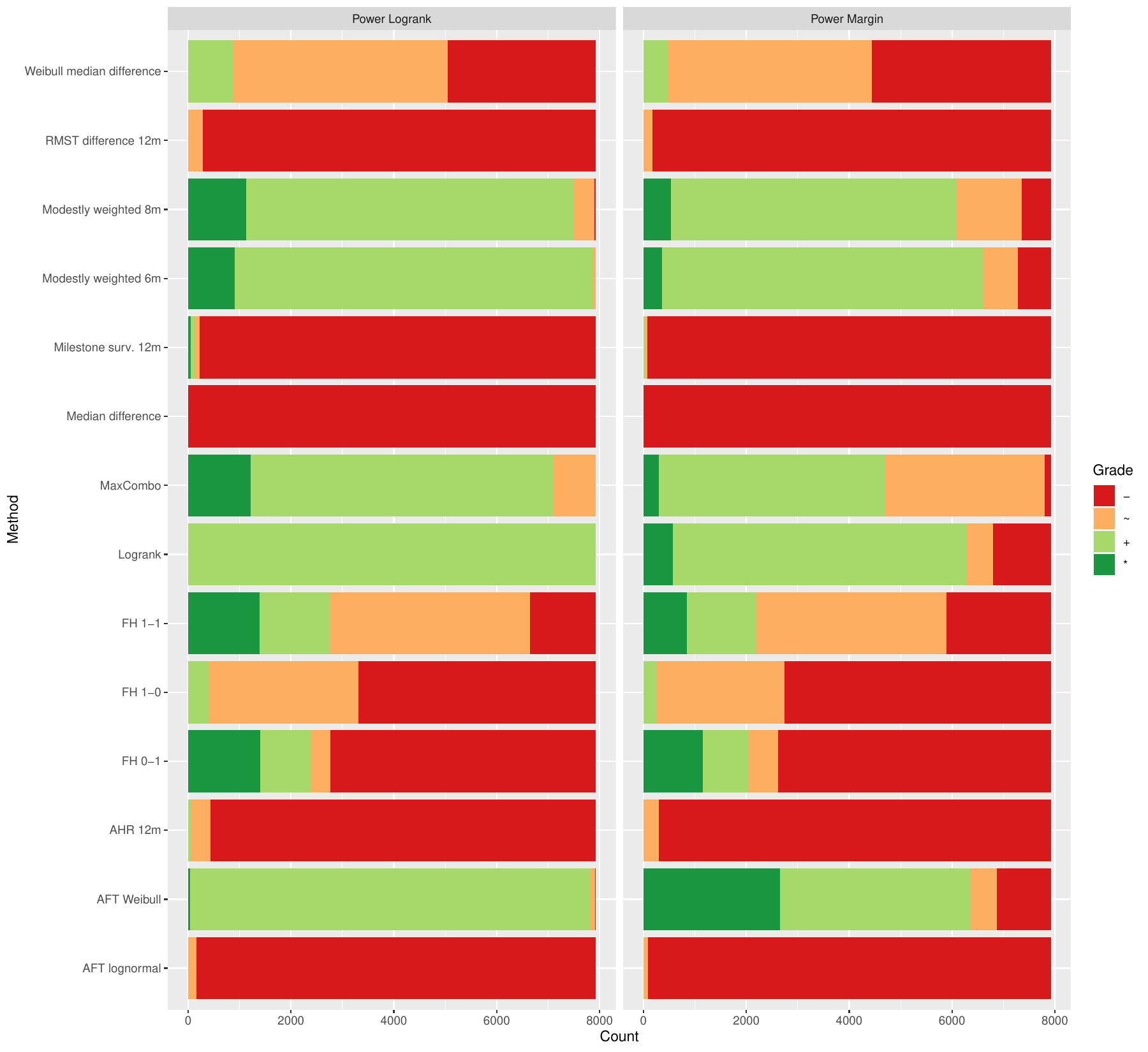} \hfill{}
\caption{Performance grades for the power of test procedures across all scenarios in the full simulation study \cite{confirms2023report}. On the horizontal axis the cumulative count of scenarios with specific performance grade is displayed. Different methods are shown on the vertical axis.}\label{fig.grades-power}
\end{figure}

Overall, in the full simulation study, in a majority of scenarios, the unweighted logrank test was among the most powerful approaches. However, in about a fifth of scenarios weighted logrank tests would result in a considerable increase in power, see Figure \ref{fig.grades-power}. 

Most studied NPH scenarios entailed increasing effects with time, hence weighted tests with emphasis on late or intermediate time-points were on average more favorable. Still, the efficiency of Fleming-Harrington weighted logrank tests was found to strongly depend on choosing a weighting function that matches the expected hazard ratio function. In contrast, the MaxCombo test was found to be robust as it was never too far from the best performing test. The modestly weighted logrank test was found to be the second most robust test after the MaxCombo test, with only few scenarios where it was considerably less powerful than the most powerful test. This is of particular interest, as the modestly weighted test has a favorable interpretation which allows to translate a difference in hazard functions to a difference in survival functions, unlike the MaxCombo or other arbitrarily weighted logrank tests.

The AFT Weibull model throughout performed similar to the logrank test. Due to its stricter model assumptions there seems no reason to employ this model as a general method of choice, though.
The studied AFT lognormal model in general had low power across different settings. Likely, the hazard functions allowed under this model do not match well with the considered scenarios and model based inference is affected by biased variance estimation. Overall, this model is not recommended in these settings.
Overall, hypothesis tests based on parameter estimates were less powerful than the (weighted) logrank tests throughout scenarios.

\begin{figure}[ht]
\includegraphics[scale=.55,clip]{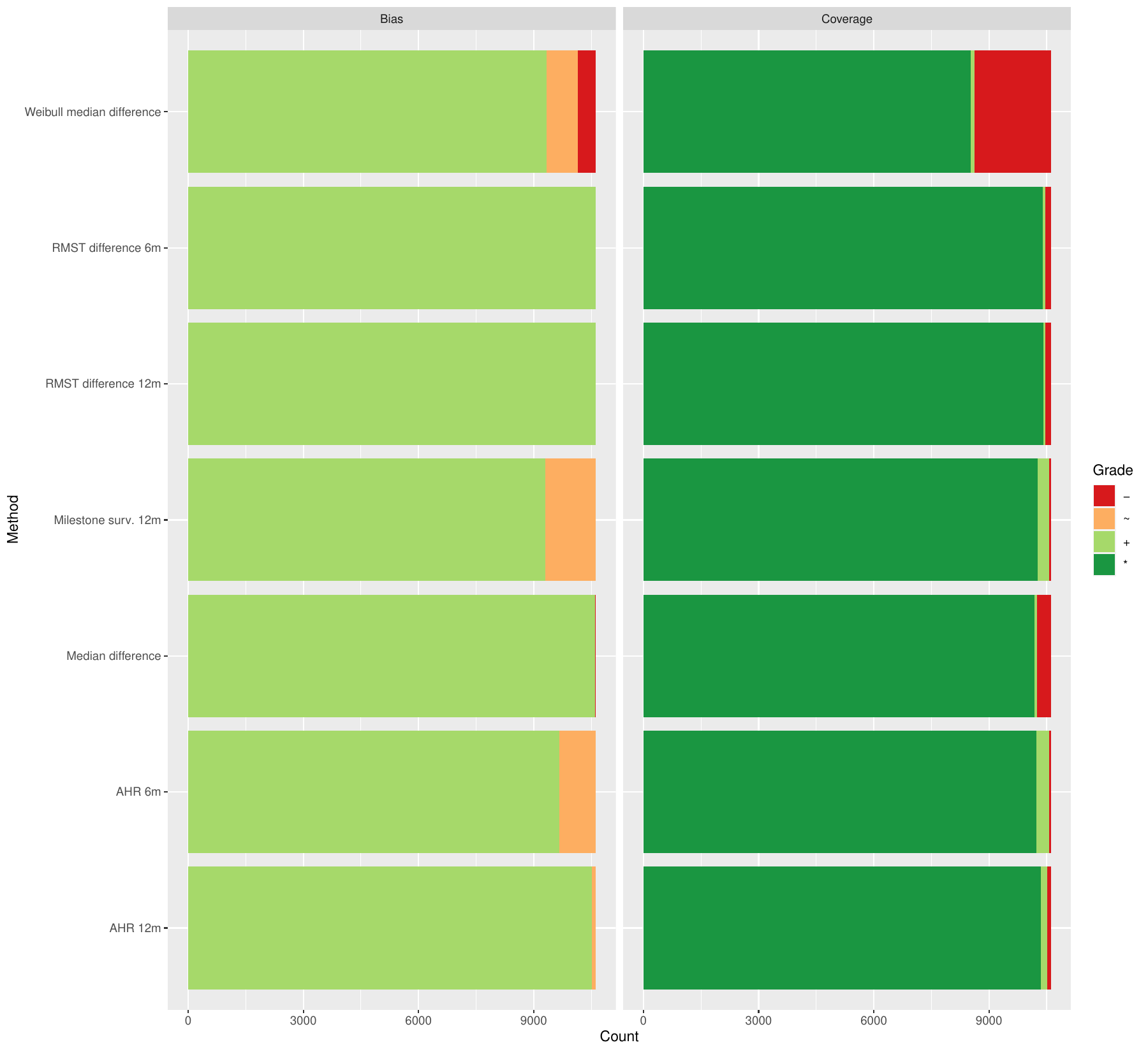} \hfill{}
\caption{Performance grades for bias and coverage of estimation methods across all scenarios in the full simulation study \cite{confirms2023report}. On the horizontal axis the cumulative count of scenarios with specific performance grade is displayed. Different methods are shown on the vertical axis.}\label{fig.grades-bias}
\end{figure}

The type I error rate was controlled, under the assumption of equal survival distributions in both groups, within acceptable limits across scenarios for all methods (right panel of Figure \ref{fig.grades-power}).
Regarding bias and confidence interval coverage, the investigated parameter estimates performed well, see Figure \ref{fig.grades-bias}. The notable exception is the median difference based on fitting a Weibull model to each group. The resulting estimate was biased in several scenarios, and in a considerable fraction of scenarios the confidence interval coverage was below the nominal level. This is likely a result of using model based inference when the model assumptions are not met.

\clearpage
\section{Case study\label{sec.case.study}}
As a case study we analysed time-to-event data that were reconstructed from
the randomised controlled phase III study ONO-4538-24 (CA209473), see the assessment report EMA/CHMP/584553/2020.
In this study, nivolumab was investigated as treatment for patients with unresectable esophageal cancer who were resistant or intolerant to standard therapy. Patients were randomized 1:1 to nivolumab or control (either docetaxel or paclitaxel). The primary endpoint was overall survival. Overall, 210 and 209 patients were randomised to nivolumab and control, respectively. The maximum follow up time was 34 months and 160 versus 173 events were observed. In the primary analysis, the groups were compared using a stratified logrank test and the hazard ratio estimate from a stratified Cox proportional hazards model. The result was a hazard ratio of 0.77 (95\% confidence interval 0.62 - 0.96) and a two-sided p-value of 0.0189.

The estimated survival curves were crossing after approximately five months at a survival probability of 0.8, with an initially better observed survival for control and eventual benefit under nivolumab after crossing. In this setting, the estimated hazard ratio from the Cox model depends on the maximum follow up time and the censoring pattern, therefore the interpretation of the reported hazard ratio is limited.

For the case study, individual participant data were reconstructed from Figure 8 of the assessment report EMA/CHMP/584553/2020 depicting the Kaplan-Meier curves of the nivolumab and control group. Of note, the subsequent analyses refer to the reconstructed data. The applied reconstruction procedure only considers the empirical survival curves for the main outcome. Information on baseline covariates, biomarkers, intercurrent events or other important outcomes were not considered. Consequently, the resulting data are not identical to the original data. Therefore, the analyses reported below, should not be misunderstood as replications of the original analysis. Any agreement or disagreement in the results should not be interpreted as confirmation or rejection of the original study results.

\begin{figure}[ht]
    \begin{center}
          \includegraphics[scale=.55,clip,trim=0cm 0cm 0.00cm 2cm]{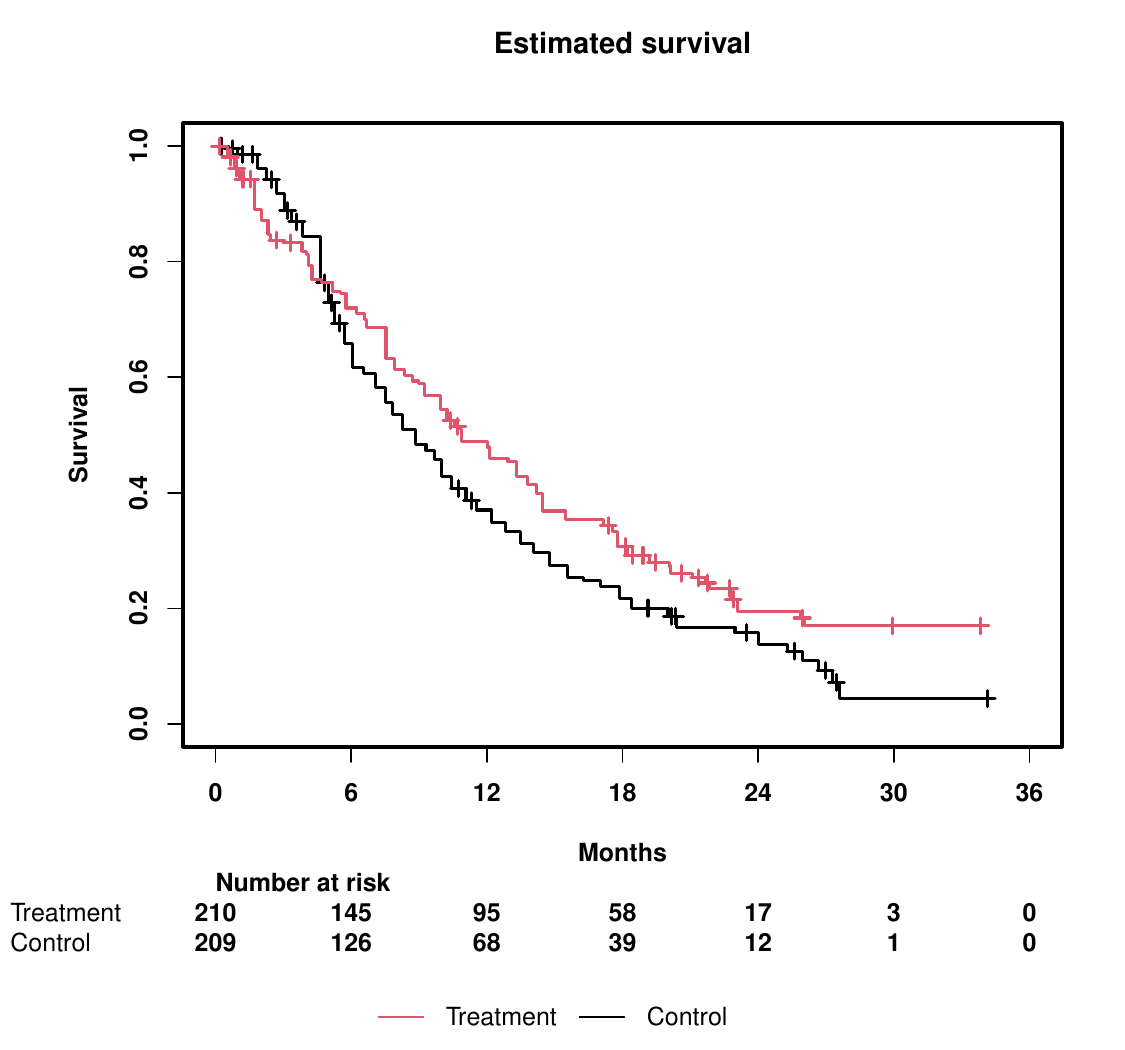}
    \end{center}
\caption{Estimated survival curves for the reconstructed data in the nivolumab case study.}
\label{fig.casestudy_surv}
\end{figure}

The estimated time-to-event distribution for both groups is shown in Figure \ref{fig.casestudy_surv} and closely matches the original figure. Crossing of the survival curves at five months is clearly visible and after crossing a steady benefit under nivolumab is observed.

\begin{figure}[ht]
    \begin{center}
          \includegraphics[scale=.55,clip,trim=0cm 0cm 0.00cm 2cm]{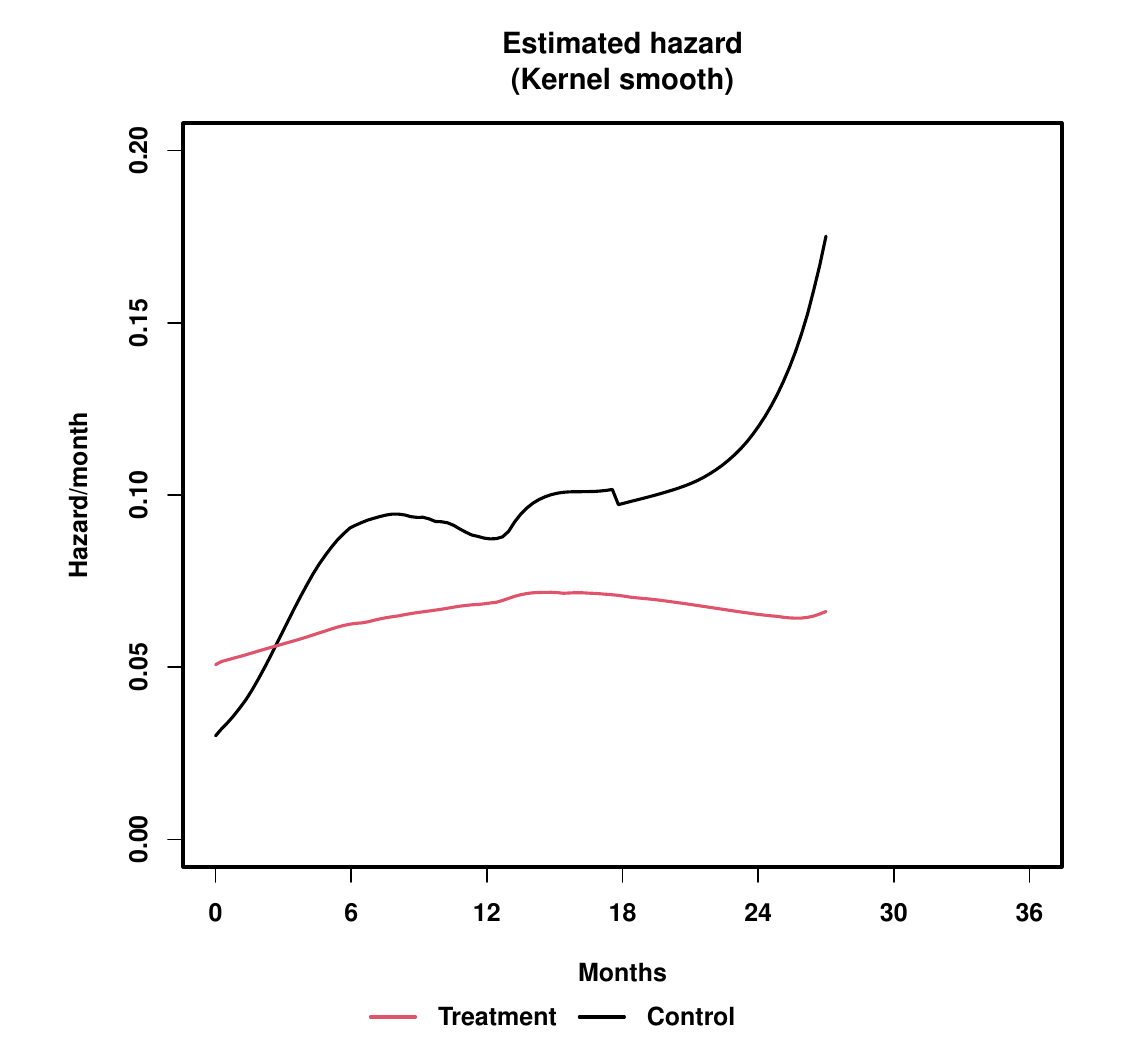}
    \end{center}
\caption{Estimated hazard functions via Kernel smoothing for the reconstructed data in the nivolumab case study.}
\label{fig.casestudy_hazard}
\end{figure}

The hazard functions in both groups were estimated using Kernel smoothing \cite{muller1994hazard}.
The resulting curves in Figure \ref{fig.casestudy_hazard} suggest that in the initial phase the hazard under control is increasing with time, surpassing the hazard under nivolumab after approximately 2.5 months, and that from month 6 on, the hazards are relatively constant in both groups. Of note, the crossing occurs at the level of the hazard at an earlier time-point than at the level of the survival function, because in the survival function, the previous reversed effect needs to be compensated over time. A benefit in the hazard, however, does not necessarily imply efficacy of the treatment, because the hazard is a local quantity and improved hazard could in principle result from a selection process at earlier times. For instance, a treatment that causes more frail patients to die early would result in a selection of fitter patients and, hence, lower hazard in the remaining population at later time points.

Different weighted logrank tests were applied to the example data set to explore the potential benefit they may have provided in the given study. These included the unweighted logrank test, the Fleming-Harrington weighted logrank tests with (rho,gamma) of (1,0), (1,1) or (0,1), the MaxCombo test comprising all four beforemenetioned tests, and the modestly weighted logrank test with with increasing weights after 6 months or after 8 months. The resulting p-values are summarized in Table \ref{tab.casestudy_pvalues} ranging from 0.005 to 0.8922.

\begin{table}[htbp]
  \centering
  \caption{Results of different weighted logrank tests in the nivolumab case study. The table shows obtained one-sided p-values for the unweighted logrank test, Fleming and Harrington (FH) rho-gamma weighted tests, the MaxCombo test comprising the unweighted logrank test and the FH weighted tests, and modestly weighted logrank tests with increasing weights after 6 or 8 months.}
    \begin{tabular}{ll}
    Method & p-value \\
    \hline
    Unweighted logrank & 0.0107 \\
    FH 1-0 & 0.0545 \\
    FH 1-1 & 0.0011 \\
    FH 0-1 & 0.0034 \\
    MaxCombo & 0.0070 \\
    Modestly weighted 6 months & 0.0051 \\
    Modestly weighted 8 months & 0.0058 \\
    \hline
    \end{tabular}
  \label{tab.casestudy_pvalues}
\end{table}

The one-sided p-value of the unweighted logrank test was 0.011, showing a significant result at the 0.025 significance level. Also, all other applied tests except for the Fleming-Harrington 1-0 weighted logrank test were significant. The 1-0 weighted test puts emphasis on early events, and as discussed, the estimated hazard was larger under nivolumab than under control for the first 2.5 months. Still, the p-value of the 1-0 weighted test was relatively small with 0.055, which suggests that even for the least optimal considered weighting, the later benefit of nivolumab was considerably.

Owing to the crossing of hazard functions, in this example the logrank test was clearly not the most significant test, but was outperformed by the 1-1 and 0-1 weighted logrank tests, the modestly weighted tests as well as the MaxCombo test. The 1-1 weighted test showed the smallest p-value of 0.001. The MaxCombo test had a notably larger p-value compared to the 1-1 and 0-1 weighted tests, probably due to including also the 1-0 weighted test. Therefore, in the given example if crossing of the hazard functions would be expected, a MaxCombo test including only the 0-0, 1-1 and 0-1 weighted tests may be a preferred approach. 
The modestly weighted tests with increasing weights after 6 months or after 8 months were similar, with p-values 0.005 and 0.006, respectively.

The two survival distributions were further compared using the difference in medians, the difference in 6 month and in 12 month survival, the difference in restricted mean survival times (RMST) up to 6 months and up to 12 months and the average hazard ratio over 6 and 12 months, all derived from the non-parametric estimate of the survival distribution. In addition, parameters derived from parametric or semi-parametric models were assessed, including the acceleration factor (AF) from a Weibull and a lognormal AFT model and the differences in medians estimated from fitting a Weibull distribution in each group. Also, the usual Cox model hazard ratio (HR) was included for comparison.

The considered parameter estimates are shown in Table \ref{tab.casestudy_estimates} with their 95\% confidence interval and a p-value for the hypothesis test of no difference between groups in the respective parameter.

\begin{table}[htbp]
  \centering
  \caption{Parameter estimates for the comparison of the two survival time distributions in the nivolumab case study.}
    \begin{tabular}{lrcr}
    Parameter & Estimate & 95\% confidence interval&     p-value\\
    \hline
 
    Survival difference 6 months & 0.06 & [-0.03, 0.15] & 0.0917 \\
    Survival difference 12 months & 0.12  & [0.02, 0.22]  & 0.0084 \\
    RMST difference 6 months & -0.18 & [-0.47, 0.11]  & 0.8922 \\
    RMST difference 12 months & 0.41  & [-0.34, 1.16]  & 0.1438 \\
    Median difference & 2.06  & [-0.75, 4.86]  & 0.0755 \\
    Weibull median difference & 1.56  & [-0.54, 3.67]  & 0.0727 \\
    Lognormal acceleration factor & 1.09  & [0.89, 1.34]  & 0.2099 \\
    Weibull acceleration factor & 1.26  & [1.06, 1.50]   & 0.0050 \\
    Cox hazard ratio & 0.77  & [0.62, 0.96]  & 0.0105 \\
    Average hazard ratio 6 months & 0.90 & [0.63, 1.29] & 0.2900 \\
    Average hazard ratio 12 months & 0.81 & [0.62, 1.06] & 0.0630 \\
    \hline
    \end{tabular}
  \label{tab.casestudy_estimates}
\end{table}

The estimated survival differences and RMST differences at 6 months and at 12 months reflect the discussed crossing of survival curves after five months. A significant survival difference of 0.12 is found at 12 months. The p-value for this 12 month survival difference is 0.0084, which is smaller than the p-value from the overall logrank test or Cox model. Hence, in the example, and in the case of crossing survival curves in general, a milestone survival analysis appears useful to determine a benefit at least for some predefined time-points. If the milestone time is well chosen, this analysis may even be more powerful than an analysis based on the logrank test or on other summary measures of the survival function.

The AFT models in this example are of limited interpretability, because regardless of the detailed distributional assumption, these models assume a monotone shift between the treatment and control survival distributions. Crossing of survival curves indicates that this assumptions is not met. Nonetheless, the acceleration factor of the Weibull model of 1.26 suggests a clear benefit of nivolumab, at least on average over the observed time frame. The lognormal model fails to detect a difference between groups.

Two separately fit Weibull distributions (as opposed to the location shift AFT Weibull model), provided an estimated median difference of 1.56 months (95\% CI -0.54 to 3.67). This was in the same range as the non-parametric estimate of the median difference of 2.06 months (95\% CI -0.75 to 4.86). With neither approach a significant benefit of nivolumab would be established, however the estimates support that the overall direction of the effect in the long term is in favor of nivolumab.

In the setting of this example study, the Cox model hazard ratio would depend strongly on the observed time span and the censoring distribution. Longer follow-up, resulting in more late events would here result in a smaller hazard ratio. The Cox model hazard ratio in the example data set was 0.77 (95\% CI 0.62 - 0.96), which is identical to the result of the original analysis.

The average hazard ratio over 6 months is close to 1, which is expected due to crossing hazards after 2.5 months. The average hazard ratio over 12 months is 0.81 indicating an increasing treatment benefit, though the corresponding hypothesis test is not yet significant with a p-value of 0.063.

Overall, a collection of predefined parameters, such as median difference and milestone survival differences may be relevant secondary outcomes in this example that help to formally establish the time-frame of treatment benefits. The semi-parametric and parametric models with one parameter for the treatment effect (Cox and AFT) models are of limited use in the quantification of the treatment effect due to obvious violation of model assumptions. More complex models that extend the Cox or AFT model could be a more appropriate choice in this example. E.g., the Yang and Prentice model allowing for the estimation of a short-term and a long-term hazard ratio could be applied \cite{Yang.2005}. Alternatively, AFT models with a distinct scale parameter for each group (using two separate Weibull distributions), or AFT models that are based on an estimated baseline hazard could be fitted \cite{pang2021flexible,crowther2023flexible}.

However, the relatively large number of events also allows for an efficient characterisation by non-parametric methods such milestone survival analyses.

\section{Discussion}\label{sec.discussion}
In the analysis of time-to-event endpoints in randomized trials, the assumption of proportional hazards provides a consistent framework of a well interpretable parameter (the hazard ratio) and corresponding efficient hypothesis tests (logrank test or asymptotically equivalent tests based on the Cox model). However, there are several scenarios that result in violation of the proportional hazards assumption, including time-dependent treatment effects, a modified treatment effect after an intercurrent event, or heterogeneity of the patient population.
Under non-proportional hazards, the power of the standard analysis approaches may be reduced, and, possibly more importantly, the interpretation of the hazard ratio estimate and corresponding hypothesis tests can be unclear. In particular, under NPH the definition of the null hypothesis for tests that are based on hazard functions need to be carefully addressed.

Especially, to assess Type I error rate control of hypothesis tests, it is important to define the null hypothesis under which type I error control is required. 
As a minimal requirement, we can demand control of the type I error rate under the null hypothesis of equal survival curves. Under this null hypothesis, type I error rate was well controlled for most considered methods in the simulation study.
However, we may also consider larger null hypotheses, e.g., including all hazard functions for which the survival curve in the treatment group lies strictly below the survival curve in the control group. Under the latter, larger null hypothesis, for example, some weighted log-rank tests may inflate the type I error rate. This may occur if very small or no weights are allocated to time points where the hazard ratio is larger than one and large weights to time points where the hazard ratio is below one. To guarantee Type I error control also under this larger null hypotheses, the modestly weighted log-rank tests have been proposed, which are robust in this respect. Also, visual inspection of the survival curves may be used to detect scenarios with obviously inferior effect in the treatment group despite a significant result of a weighted logrank test. A formal combination of weighted logrank tests and decision criteria based on estimated survival curves is a field of further research. Of note, type I error control under null hypotheses more general than equal survival curves was not systematically assessed in the simulation study.

As expected, in settings where the proportional hazards assumption was substantially violated, methods tailored to the specific type of deviation (as the Fleming-Harrington tests with appropriate weights, the MaxCombo test or the modestly weighted tests) had a larger power compared to methods that have the largest power under proportional hazards as the logrank test or the Cox-model. 
The logrank test, the Cox model Wald test and the AFT Weibull Wald test had similar power across the considered scenarios. This is expected, as the logrank and Cox Wald tests are asymptotically equivalent and the Weibull model is also a PH model and allows for decreasing or increasing baseline hazard functions.
However, as the Weibull baseline hazard will likely not match the actual hazard function, model based inference from this model is not guaranteed to result in nominal coverage of confidence intervals or unbiased estimation of derived parameters such as median survival times. Accordingly, deviations in confidence interval coverage from the nominal level and biased medians (estimated from two Weibull distributions separately fit to the treatment groups) were observed in the simulation study.

Across settings, the power of the MaxCombo test was typically close to or above the power of the logrank test. It can therefore be considered a robust method. However, the results of weighted logrank tests in general are difficult to interpret beyond the conclusion that survival curves are not equal. It is not obvious which time points contributed how strongly to the final test decision, and this difficulty is aggravated for the MaxCombo test. Also a local difference between hazard functions may result from selection effects in a population with heterogeneously frail patients and may not be sufficient to conclude a treatment benefit \cite{aalen1994effects,hernan2010hazards}. 
Notably, selection in favour of the experimental treatment can occur only if the control was superior compared to treatment in some earlier time interval.
 
Under delayed onset and crossing hazards, the modestly weighted tests typically had less power than a test more decisively weighted for late events, but the more robust interpretation of the modestly weighted test in terms of directly translating to a beneficial difference in survival function (even if only at least at one time point) makes these tests a viable choice. Under proportional hazards and under reduced effects with time, the modestly weighted test showed power similar to the unweighted logrank test and therefore may be considered to be a robust choice. 

Overall, the simulation studies confirmed that the weighted logrank tests and the MaxCombo test can have a larger power than the unweighted logrank test, especially when there is a strong deviation from the proportional hazards assumption and a defined weight function emphasises effects at the corresponding event times. This comes at the cost of a somewhat lower power when the proportional hazards assumption holds and that the tests may have a larger rejection probability in scenarios where the survival curves differ but the treatment is not considered to have a favourable survival distribution. 

Hypothesis tests based on statistics that cover only a part of the survival or hazard function up to a specified time-point, such as differences in RMST, the considered concordance-type average hazard ratio, median survival times or milestone survival probabilities, had in general lower power than the logrank tests. However, these tests test null hypotheses related to well interpretable parameters and control the type I error rate under the broader null hypotheses that these parameters are equal, rather than just the null hypothesis of equality of the survival curves. In contrast, e.g., for the logrank tests the rejection probability may be larger than the nominal under scenarios of equal RMSTs, medians or milestone survival probabilities.

The lower power of the considered estimate-based tests may be expected in scenarios where the treatment effect is stronger at later time-points, which are in part not covered by these statistics. However, also in scenarios with stronger early effects, the effective sample size for statistical inference with these statistics is smaller than for statistics that cover all observed event times, which may further explain their lower power in many settings. 

However, under non-proportional hazards scenarios are possible, where there is a large differences in the survival function at particular time points only. In such cases, tests for milestone survival differences (at suitable chosen milestone times) can have larger power than a logrank test. A possible example is the nivolumab case study, where the difference in 12 month survival probabilities resulted in a smaller p-value than the logrank test. However, the Fleming-Harrington 1-1 and 0-1 weighted logrank test, the MaxCombo test and the modestly weighted logrank test all had lower p-values.

Regarding the direct comparison of RMST difference to the unweighted logrank test, Tian et al. \cite{tian2018efficiency} studied the relative asymptotic efficiency. They found that for many scenarios with small effect at early time points and larger effect at later time-points, the logrank test was more efficient than the RMST difference. In other scenarios, however, RMST can have a larger efficiency than the logrank test. 

The asymptotic normal approximation for the studied estimates, that is used in the considered hypothesis tests and confidence intervals, may be insufficient if the number at risk at relevant time-points is small, which may be an issue in particular for median survival and milestone survival probabilities. In the simulation study, bias was in general small for estimates of the difference in median survival, milestone survival probabilities and RMSTs that were derived from the non-parametrically estimated survival functions. Confidence interval coverage was close to the nominal level for RMST and milestone survival differences. For median survival differences, confidence interval coverage was in general close to the nominal level with the exception of a few scenarios where coverage probabilities a few percentage points below the nominal coverage were observed.

Estimates from model based methods such as the AFT Weibull and the AFT lognormal model may give biased results if the true hazard functions do not meet the model assumptions. Accordingly, non-negligible bias and confidence interval coverage below the nominal level was observed for median differences calculated from Weibull models in several scenarios. Due to their strong model assumptions and non-robustness of the obtained estimates with respect to violations of these assumptions, the studied standard AFT models cannot be considered to be a robust class of methods in the studied NPH scenarios. However, AFT models with relaxed distributional assumptions have been suggested, that, for example, include time-dependent acceleration factors \cite{pang2021flexible,crowther2023flexible}. These models can provide an improved fit.

In the planning of confirmatory clinical trials, the definition of an appropriate estimand is an prerequisite for an unambiguous interpretation of trial results \cite{E9addendumestimands,unkel2019estimands}. The estimand framework includes defining the target population, the treatment conditions, intercurrent events, the outcome variable and a population level summary measure to quantify differences between treatment conditions with respect to the outcome variable. NPH will usually not affect the definition of the population, treatment conditions or intercurrent events, nor the importance of survival time or other time-to-event variables as primary outcome. However, under NPH, the definition of a summary measure is more challenging. As noted above, the hazard ratio is not a suitable summary measure under NPH. Parameters such as differences in RMST or milestone survival probabilities are an alternative which are well defined also under NPH. However, the potential lack in power of corresponding hypothesis tests is a disadvantage of these measures. The often more powerful (weighted) log rank tests, however, lack an easily interpretable summary measure and do not provide a quantification of treatment effects for benefit risk assessment. 

Understanding the composition of the study population, the mode of action of the treatments under investigation, and the specifics of the trial design are important to identify reasons for non-proportional hazards. For instance, if the mode of action suggests a delayed onset of treatment or differential effects in subgroups, this information can guide the formulation of an analysis strategy that directly accounts for factors causing non-proportional hazards in the entire population. Consider, for example, the IPASS study, which compared gefitinib to carboplatin–paclitaxel in pulmonary adenocarcinoma \cite{mok2009gefitinib}. This study presented crossing survival curves for the primary progression-free survival endpoint. Yet, a subgroup analysis focusing on EGFR-mutation revealed distinct survival curves for EGFR positive and EGFR negative patients. These curves displayed opposing treatment effects but no indication of non-proportional hazards. Thus, in the latter case a stratified analysis might be a more suitable choice than dedicated statistical methods, targeting non-proportional hazards.

As a limitation of this study, being inherent to all simulation studies, it cannot possibly cover the whole parameter space and all relevant data generating models. 
For example, informative censoring was not investigated in this study and is a topic for further research. Informative censoring may introduce bias for all considered methods. The amount of bias may depend on the data generating mechanism as well as the testing and estimation methods considered.
Nevertheless, a large range of scenarios for confirmatory trials has been covered, as we used diverse approaches to generate the data, including delayed onsets of treatment effects, crossing hazards, a multi-state model (progression – death) and subgroups. A further limitation concerns  the summary of results based on the grading of methods. This summary is specific to the choice of simulation scenarios and the grading scheme.
\\ \ \\

\noindent\textbf{Conclusions}

\noindent Under non proportional hazards there are many options to assess the difference in survival functions and a single parameter cannot quantify this difference comprehensively. Appropriate summary statistics and hypothesis tests need to be chosen taking into account the expected characteristics of the survival functions, such as the timing of possible delayed onsets or phases of waning treatment effect. 
Under NPH, there is a trade off between interpretability and power when choosing an analysis strategy. 

Weighted logrank tests may have larger power than the unweighted logrank test under NPH, and also typically provide larger power than tests based on estimates. However, 
when weighted tests are applied, it must be shown that a significant result indeed corresponds to a treatment benefit in terms of relevant parameters. This may either limit the use of weighted logrank tests to those that test the larger null hypothesis of dominating survival curves (unweighted or modestly weighted test) or it requires additional assumptions, e.g. assuming absence of detrimental treatment effects.
The RMST difference, the average hazard ratio, differences in milestone survival or differences in quantiles, such as the median survival time, may serve as scalar effect measures and could be considered to replace the Cox model hazard ratio under NPH. However, their interpretation depends on the defined time-points or quantiles and their power is often lower than the power of the standard logrank test. 

The interpretation of empirical survival curves may aid in the discussion of results. Reporting several pre-defined summary measures can help to contextualise the results and facilitate the interpretation of treatment effects \cite{ristl2023simultaneous}. However, to draw robust conclusions based on multiple parameters or the inspection of the whole empirical survival curves, the variability of these estimates needs to be taken into account.

\section*{Funding}
This work has received funding from the European Medicines Agency (Re-opening of competition \newline EMA/2020/46/TDA/L3.02 (Lot 3))

\section*{Acknowledgement}
The authors would like to thank Juan José Abellán and Marcia Rückbeil of the European Medicines Agency as well as Andreas Brandt of the Bundesinstitut für Arzneimittel und Medizinprodukte for valuable comments and insightful discussions. This document expresses the opinion of the authors of the paper, and may not be understood or quoted as being made on behalf of or reflecting the position of the European Medicines Agency or one of its committees or working parties.

\bibliographystyle{unsrt}
\bibliography{bibliography}

\begin{thebibliography}{10}

\bibitem{peto1972asymptotically}
Richard Peto and Julian Peto.
\newblock Asymptotically efficient rank invariant test procedures.
\newblock {\em Journal of the Royal Statistical Society: Series A (General)}, 135(2):185--198, 1972.

\bibitem{anagnostou2017immuno}
Valsamo Anagnostou, Mark Yarchoan, Aaron~R Hansen, Hao Wang, Franco Verde, Elad Sharon, Deborah Collyar, Laura~QM Chow, and Patrick~M Forde.
\newblock Immuno-oncology trial endpoints: capturing clinically meaningful activity.
\newblock {\em Clinical Cancer Research}, 23(17):4959--4969, 2017.

\bibitem{shen2023nonproportional}
Yuan-Li Shen, Xin Wang, Mushti Sirisha, Flora Mulkey, Jiaxi Zhou, Xin Gao, Lijun Zhang, Thomas Gwise, Shenghui Tang, Marc Theoret, et~al.
\newblock Nonproportional hazards—an evaluation of the maxcombo test in cancer clinical trials.
\newblock {\em Statistics in Biopharmaceutical Research}, 15(2):300--309, 2023.

\bibitem{aalen1994effects}
Odd~O Aalen.
\newblock Effects of frailty in survival analysis.
\newblock {\em Statistical methods in medical research}, 3(3):227--243, 1994.

\bibitem{xu2000estimating}
Ronghui Xu and John O’Quigley.
\newblock Estimating average regression effect under non-proportional hazards.
\newblock {\em Biostatistics}, 1(4):423--439, 2000.

\bibitem{hernan2010hazards}
Miguel~A Hern{\'a}n.
\newblock The hazards of hazard ratios.
\newblock {\em Epidemiology (Cambridge, Mass.)}, 21(1):13, 2010.

\bibitem{martinussen2020subtleties}
Torben Martinussen, Stijn Vansteelandt, and Per~Kragh Andersen.
\newblock Subtleties in the interpretation of hazard contrasts.
\newblock {\em Lifetime Data Analysis}, 26:833--855, 2020.

\bibitem{aalen2015does}
Odd~O Aalen, Richard~J Cook, and Kjetil R{\o}ysland.
\newblock Does cox analysis of a randomized survival study yield a causal treatment effect?
\newblock {\em Lifetime Data Analysis}, 21:579--593, 2015.

\bibitem{posch2022testing}
Martin Posch, Robin Ristl, and Franz K{\"o}nig.
\newblock Testing and interpreting the “right” hypothesis—comment on “non-proportional hazards—an evaluation of the maxcombo test in cancer clinical trials”.
\newblock {\em Statistics in Biopharmaceutical Research}, pages 1--2, 2022.

\bibitem{bartlett2020hazards}
Jonathan~W Bartlett, Tim~P Morris, Mats~J Stensrud, Rhian~M Daniel, Stijn~K Vansteelandt, and Carl-Fredrik Burman.
\newblock The hazards of period specific and weighted hazard ratios.
\newblock {\em Statistics in Biopharmaceutical Research}, 12(4):518, 2020.

\bibitem{magirr2019modestly}
Dominic Magirr and Carl-Fredrik Burman.
\newblock Modestly weighted logrank tests.
\newblock {\em Statistics in medicine}, 38(20):3782--3790, 2019.

\bibitem{bardo2023methods}
Maximilian Bardo, Cynthia Huber, Norbert Benda, Jonas Brugger, Tobias Fellinger, Vaidotas Galaune, Judith Heinz, Harald Heinzl, Andrew~C Hooker, Florian Klinglm{\"u}ller, et~al.
\newblock Methods for non-proportional hazards in clinical trials: A systematic review.
\newblock {\em arXiv preprint arXiv:2306.16858}, 2023.

\bibitem{fleming1981class}
Thomas~R Fleming and David~P Harrington.
\newblock A class of hypothesis tests for one and two sample censored survival data.
\newblock {\em Communications in Statistics-Theory and Methods}, 10(8):763--794, 1981.

\bibitem{zucker1990weighted}
David~M Zucker and Edward Lakatos.
\newblock Weighted log rank type statistics for comparing survival curves when there is a time lag in the effectiveness of treatment.
\newblock {\em Biometrika}, 77(4):853--864, 1990.

\bibitem{jimenez2019properties}
Jos{\'e}~L Jim{\'e}nez, Viktoriya Stalbovskaya, and Byron Jones.
\newblock Properties of the weighted log-rank test in the design of confirmatory studies with delayed effects.
\newblock {\em Pharmaceutical statistics}, 18(3):287--303, 2019.

\bibitem{liu2018weighted}
Shufang Liu, Chenghao Chu, and Alan Rong.
\newblock Weighted log-rank test for time-to-event data in immunotherapy trials with random delayed treatment effect and cure rate.
\newblock {\em Pharmaceutical statistics}, 17(5):541--554, 2018.

\bibitem{lee2007versatility}
Seung-Hwan Lee.
\newblock On the versatility of the combination of the weighted log-rank statistics.
\newblock {\em Computational statistics \& data analysis}, 51(12):6557--6564, 2007.

\bibitem{fleming2011counting}
Thomas~R Fleming and David~P Harrington.
\newblock {\em Counting processes and survival analysis}, volume 169.
\newblock John Wiley \& Sons, 2011.

\bibitem{lin2020alternative}
Ray~S Lin, Ji~Lin, Satrajit Roychoudhury, Keaven~M Anderson, Tianle Hu, Bo~Huang, Larry~F Leon, Jason~JZ Liao, Rong Liu, Xiaodong Luo, et~al.
\newblock Alternative analysis methods for time to event endpoints under nonproportional hazards: a comparative analysis.
\newblock {\em Statistics in Biopharmaceutical Research}, 12(2):187--198, 2020.

\bibitem{ristl2021delayed}
Robin Ristl, Nicol{\'a}s~M Ballarini, Heiko G{\"o}tte, Armin Sch{\"u}ler, Martin Posch, and Franz K{\"o}nig.
\newblock Delayed treatment effects, treatment switching and heterogeneous patient populations: How to design and analyze {RCTs} in oncology.
\newblock {\em Pharmaceutical statistics}, 20(1):129--145, 2021.

\bibitem{royston2011use}
Patrick Royston and Mahesh~KB Parmar.
\newblock The use of restricted mean survival time to estimate the treatment effect in randomized clinical trials when the proportional hazards assumption is in doubt.
\newblock {\em Statistics in medicine}, 30(19):2409--2421, 2011.

\bibitem{rauch2018average}
Geraldine Rauch, Werner Brannath, Matthias Brueckner, and Meinhard Kieser.
\newblock The average hazard ratio--a good effect measure for time-to-event endpoints when the proportional hazard assumption is violated?
\newblock {\em Methods of information in medicine}, 57(03):089--100, 2018.

\bibitem{shafrin2017patient}
Jason Shafrin, Taylor~T Schwartz, Tony Okoro, and John~A Romley.
\newblock Patient versus physician valuation of durable survival gains: implications for value framework assessments.
\newblock {\em Value in Health}, 20(2):217--223, 2017.

\bibitem{freidlin2019methods}
Boris Freidlin and Edward~L Korn.
\newblock Methods for accommodating nonproportional hazards in clinical trials: ready for the primary analysis?
\newblock {\em Journal of Clinical Oncology}, 37(35):3455, 2019.

\bibitem{li2015statistical}
Huimin Li, Dong Han, Yawen Hou, Huilin Chen, and Zheng Chen.
\newblock Statistical inference methods for two crossing survival curves: a comparison of methods.
\newblock {\em PLoS One}, 10(1):e0116774, 2015.

\bibitem{ananthakrishnan2021critical}
Revathi Ananthakrishnan, Stephanie Green, Alessandro Previtali, Rong Liu, Daniel Li, and Michael LaValley.
\newblock Critical review of oncology clinical trial design under non-proportional hazards.
\newblock {\em Critical Reviews in Oncology/Hematology}, 162:103350, 2021.

\bibitem{benda2010aspects}
Norbert Benda, Michael Branson, Willi Maurer, and Tim Friede.
\newblock Aspects of modernizing drug development using clinical scenario planning and evaluation.
\newblock {\em Drug information journal: DIJ/Drug Information Association}, 44:299--315, 2010.

\bibitem{friede2010refinement}
Tim Friede, Richard Nicholas, Nigel Stallard, Susan Todd, Nicholas Parsons, Elsa Vald{\'e}s-M{\'a}rquez, and Jeremy Chataway.
\newblock Refinement of the clinical scenario evaluation framework for assessment of competing development strategies with an application to multiple sclerosis.
\newblock {\em Drug information journal: DIJ/Drug Information Association}, 44:713--718, 2010.

\bibitem{morris2019using}
Tim~P Morris, Ian~R White, and Michael~J Crowther.
\newblock Using simulation studies to evaluate statistical methods.
\newblock {\em Statistics in medicine}, 38(11):2074--2102, 2019.

\bibitem{Aalen.2008}
Odd~O. Aalen, {\O}rnulf Borgan, and H{\aa}kon~K. Gjessing.
\newblock {\em Event History Analysis: A Process Point of View}.
\newblock Statistics for Biology and Health. Springer, Dordrecht, 2008.

\bibitem{aalen2010history}
Odd~O Aalen, Per~Kragh Andersen, {\O}rnulf Borgan, Richard~D Gill, and Niels Keiding.
\newblock History of applications of martingales in survival analysis.
\newblock {\em arXiv preprint arXiv:1003.0188}, 2010.

\bibitem{harrington1982class}
David~P Harrington and Thomas~R Fleming.
\newblock A class of rank test procedures for censored survival data.
\newblock {\em Biometrika}, 69(3):553--566, 1982.

\bibitem{fleming1991counting}
Thomas~R Fleming and David~P Harrington.
\newblock {\em Counting processes and survival analysis}.
\newblock John Wiley \& Sons, 1991.

\bibitem{tarone1981distribution}
Robert~E Tarone.
\newblock On the distribution of the maximum of the logrank statistic and the modified wilcoxon statistic.
\newblock {\em Biometrics}, pages 79--85, 1981.

\bibitem{karrison2016versatile}
Theodore~G Karrison et~al.
\newblock Versatile tests for comparing survival curves based on weighted log-rank statistics.
\newblock {\em Stata Journal}, 16(3):678--690, 2016.

\bibitem{hasegawa2020restricted}
Takahiro Hasegawa, Saori Misawa, Shintaro Nakagawa, Shinichi Tanaka, Takanori Tanase, Hiroyuki Ugai, Akira Wakana, Yasuhide Yodo, Satoru Tsuchiya, Hideki Suganami, et~al.
\newblock Restricted mean survival time as a summary measure of time-to-event outcome.
\newblock {\em Pharmaceutical statistics}, 19(4):436--453, 2020.

\bibitem{ristl2023simultaneous}
Robin Ristl, Heiko G{\"o}tte, Armin Sch{\"u}ler, Martin Posch, and Franz K{\"o}nig.
\newblock Simultaneous inference procedures for the comparison of multiple characteristics of two survival functions.
\newblock {\em arXiv:2310.01990}, 2023.

\bibitem{klein2003survival}
John~P Klein, Melvin~L Moeschberger, et~al.
\newblock {\em Survival analysis: Techniques for censored and truncated data, second edition}.
\newblock Springer, 2003.

\bibitem{kalbfleisch1981estimation}
John~D Kalbfleisch and Ross~L Prentice.
\newblock Estimation of the average hazard ratio.
\newblock {\em Biometrika}, 68(1):105--112, 1981.

\bibitem{confirms2023report}
{CONFIRMS}.
\newblock A simulation study to evaluate the performance characteristics of statistical methods for the analysis of time-to-event data under non-proportional hazards.
\newblock 2023.

\bibitem{schoenfeld1981asymptotic}
David Schoenfeld.
\newblock The asymptotic properties of nonparametric tests for comparing survival distributions.
\newblock {\em Biometrika}, 68(1):316--319, 1981.

\bibitem{muller1994hazard}
Hans-Georg Muller and Jane-Ling Wang.
\newblock Hazard rate estimation under random censoring with varying kernels and bandwidths.
\newblock {\em Biometrics}, pages 61--76, 1994.

\bibitem{Yang.2005}
Song Yang and Ross Prentice.
\newblock Semiparametric analysis of short-term and long-term hazard ratios with two-sample survival data.
\newblock {\em Biometrika}, 92(1):1--17, 2005.

\bibitem{pang2021flexible}
Menglan Pang, Robert~W Platt, Tibor Schuster, and Michal Abrahamowicz.
\newblock Flexible extension of the accelerated failure time model to account for nonlinear and time-dependent effects of covariates on the hazard.
\newblock {\em Statistical Methods in Medical Research}, 30(11):2526--2542, 2021.

\bibitem{crowther2023flexible}
Michael~J Crowther, Patrick Royston, and Mark Clements.
\newblock A flexible parametric accelerated failure time model and the extension to time-dependent acceleration factors.
\newblock {\em Biostatistics}, 24(3):811--831, 2023.

\bibitem{tian2018efficiency}
Lu~Tian, Haoda Fu, Stephen~J Ruberg, Hajime Uno, and Lee-Jen Wei.
\newblock Efficiency of two sample tests via the restricted mean survival time for analyzing event time observations.
\newblock {\em Biometrics}, 74(2):694--702, 2018.

\bibitem{E9addendumestimands}
{Committee for Medicinal Products for Human Use}.
\newblock {ICH E9 (R1) addendum on estimands and sensitivity analysis in clinical trials to the guideline on statistical principles for clinical trials}.
\newblock {\em EMA/CHMP/ICH/436221/2017}, 2020.

\bibitem{unkel2019estimands}
Steffen Unkel, Marjan Amiri, Norbert Benda, Jan Beyersmann, Dietrich Knoerzer, Katrin Kupas, Frank Langer, Friedhelm Leverkus, Anja Loos, Claudia Ose, et~al.
\newblock On estimands and the analysis of adverse events in the presence of varying follow-up times within the benefit assessment of therapies.
\newblock {\em Pharmaceutical Statistics}, 18(2):166--183, 2019.

\bibitem{mok2009gefitinib}
Tony~S Mok, Yi-Long Wu, Sumitra Thongprasert, Chih-Hsin Yang, Da-Tong Chu, Nagahiro Saijo, Patrapim Sunpaweravong, Baohui Han, Benjamin Margono, Yukito Ichinose, et~al.
\newblock Gefitinib or carboplatin--paclitaxel in pulmonary adenocarcinoma.
\newblock {\em New England Journal of Medicine}, 361(10):947--957, 2009.

\end{thebibliography}

\clearpage
\section*{Appendix - Supplementary Material}
\renewcommand{\thefigure}{S\arabic{figure}}
\setcounter{figure}{0}
\renewcommand{\thetable}{S\arabic{table}}
\setcounter{table}{0}

The Supplementary Material shows the survival functions, hazard functions and hazard ratio functions for all simulation scenarios considered in the paper. For survival functions and hazard functions, black lines correspond to the control group and green lines correspond to the treatment group. 

Additionally the distribution of the follow-up time across simulation runs is shown in terms of the mean follow-up time, represented by a vertical solid line, and mean +/- 2 standard deviations, represented by vertical dashed lines.

Details regarding the parameter values on the simulation scenarios are found in Section \ref{sec.simulation.scenarios} of the main paper.

\begin{figure}[!htbp]
\begin{center}
\includegraphics[page=1,scale=.7,clip,trim=0cm 0cm 0.00cm 0cm]{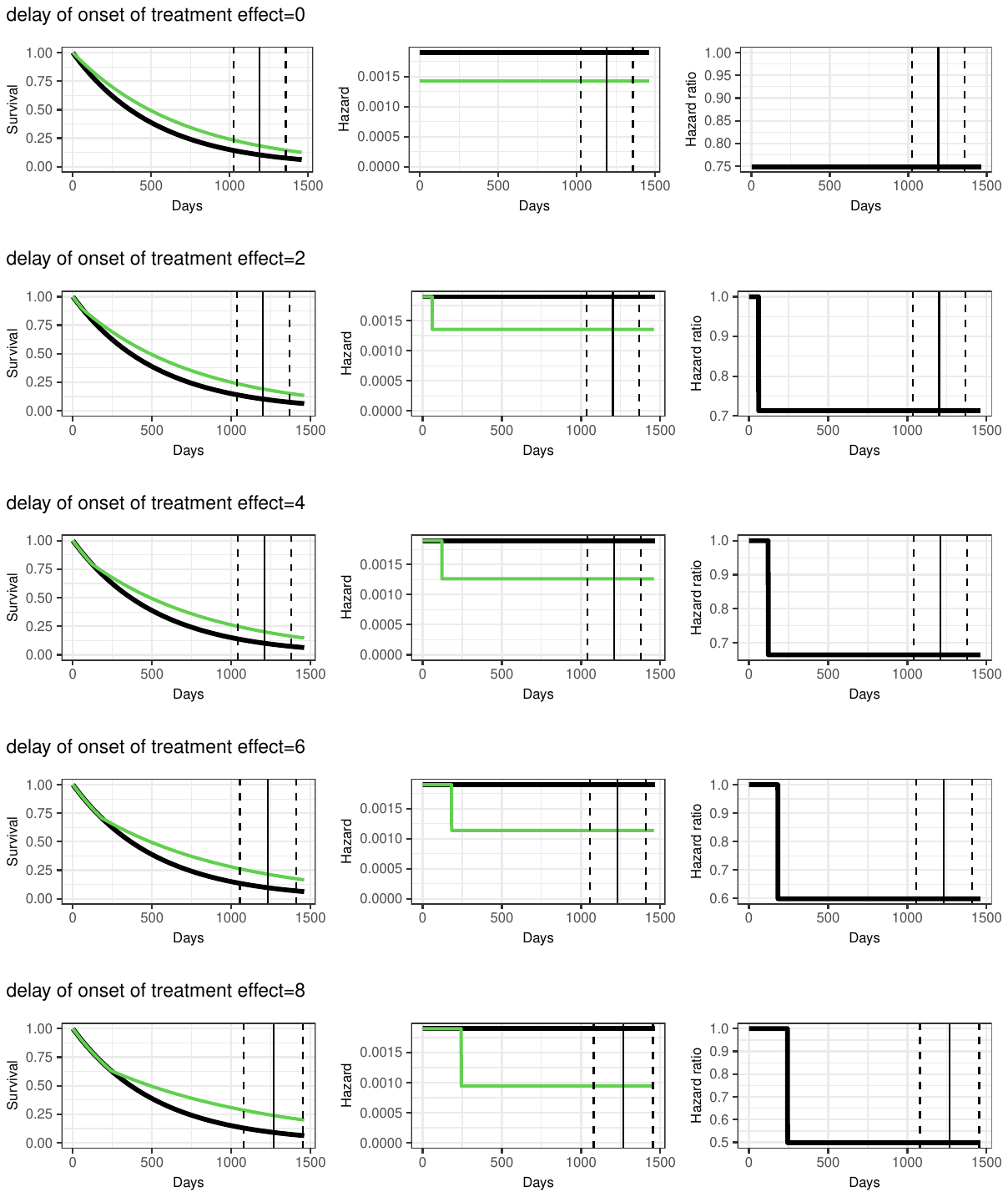}
\end{center}
\caption{Survival functions, hazard functions and hazard ratio functions in the scenarios with delayed onset of treatment effect. Each row of the figure represents a scenario with a specific delay time as indicated in the superscripts. The left panel shows survival functions for treatment (green) and control (black). In each row, the center panel shows the corresponding hazard functions. The right panel shoes the hazard ratio function. Additionally, the distribution of the follow-up time across simulation runs is shown in terms of the mean follow-up time, represented by a vertical solid line, and mean +/- 2 standard deviations, represented by vertical dashed lines.}
\label{fig.scenario_delay}
\end{figure}

\begin{figure}[!htbp]
\begin{center}
\includegraphics[page=2,scale=.7,clip,trim=0cm 0cm 0.00cm 0cm]{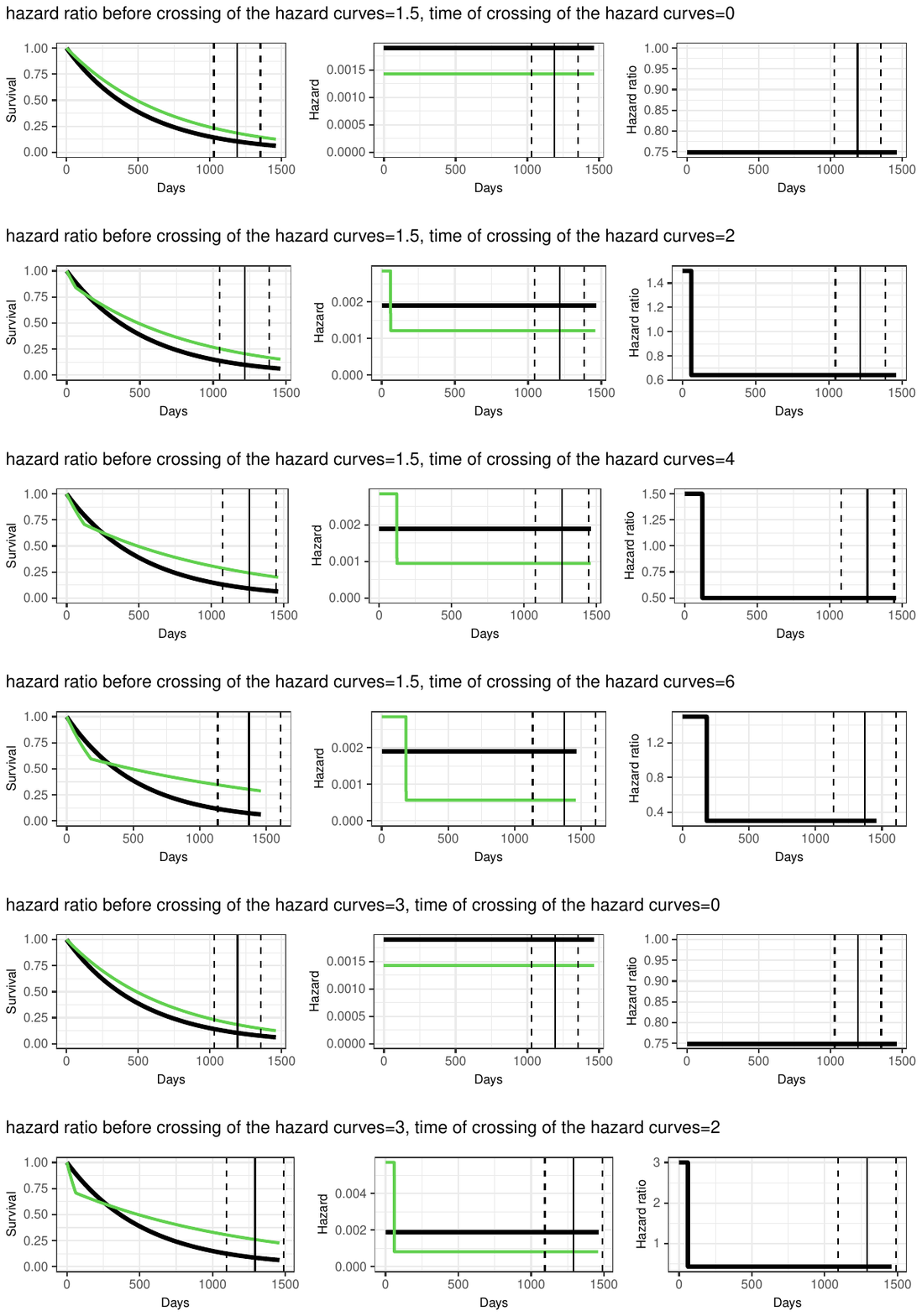}
\end{center}
\caption{Survival functions, hazard functions and hazard ratio functions in the scenarios with crossing hazards. Each row of the figure represents a scenario with a specific crossing time as indicated in the superscripts. The left panel shows survival functions for treatment (green) and control (black). In each row, the center panel shows the corresponding hazard functions. The right panel shoes the hazard ratio function. Additionally, the distribution of the follow-up time across simulation runs is shown in terms of the mean follow-up time, represented by a vertical solid line, and mean +/- 2 standard deviations, represented by vertical dashed lines.}
\label{fig.scenario_crossing}
\end{figure}

\begin{figure}[!htbp]
\begin{center}
\includegraphics[page=3,scale=.7,clip,trim=0cm 0cm 0.00cm 0cm]{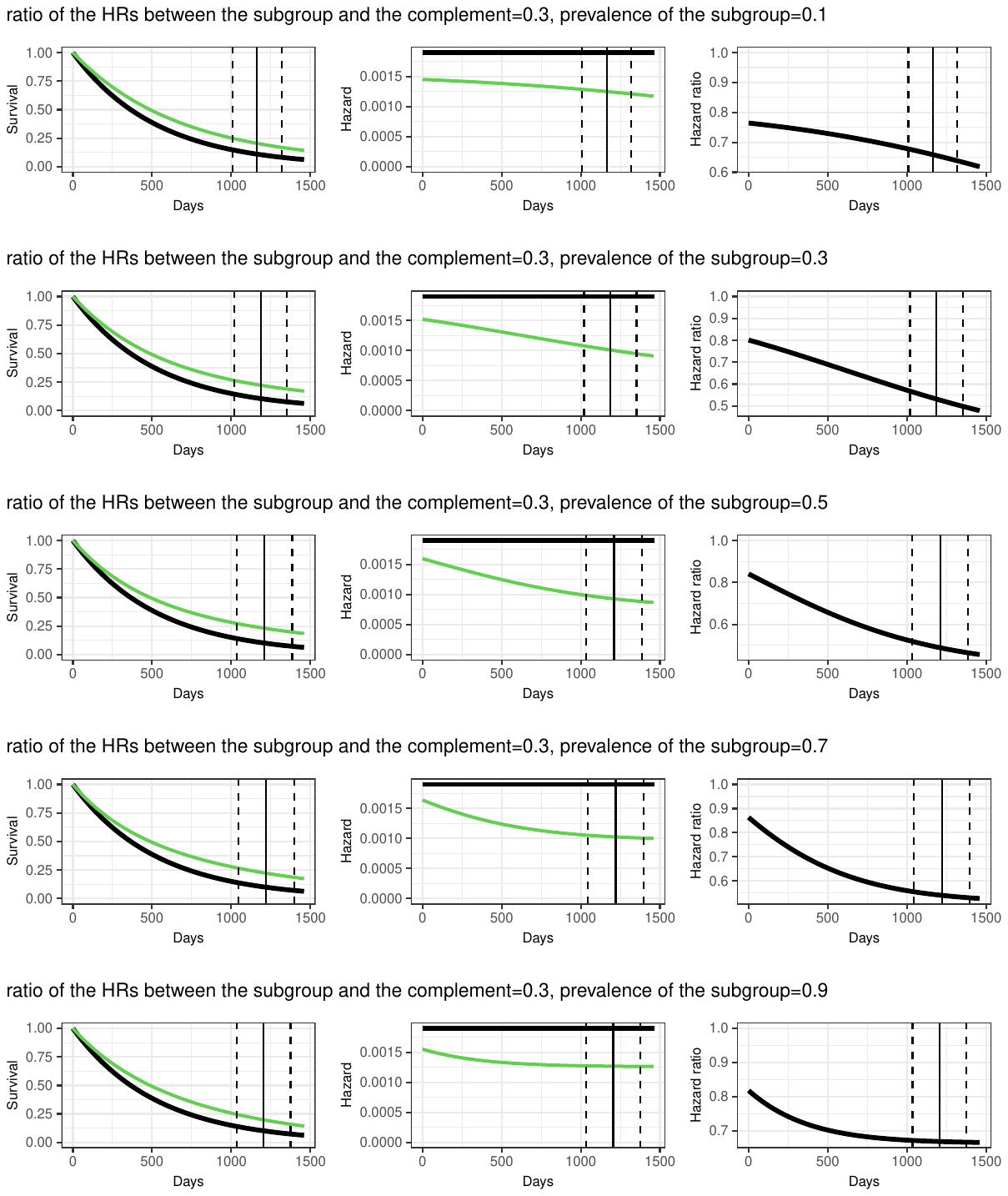}
\end{center}
\caption{Survival functions, hazard functions and hazard ratio functions in the scenarios with biomarker subgroups. Each row of the figure represents a scenario with a specific hazard ratio between biomarker positive and negative patients and specific prevalence of biomarker positive patients as indicated in the superscripts. The left panel shows survival functions for treatment (green) and control (black). In each row, the center panel shows the corresponding hazard functions. The right panel shoes the hazard ratio function. Additionally, the distribution of the follow-up time across simulation runs is shown in terms of the mean follow-up time, represented by a vertical solid line, and mean +/- 2 standard deviations, represented by vertical dashed lines.}
\label{fig.scenario_subgroups}
\end{figure}

\begin{figure}[!htbp]
\begin{center}
\includegraphics[page=4,scale=.7,clip,trim=0cm 0cm 0.00cm 0cm]{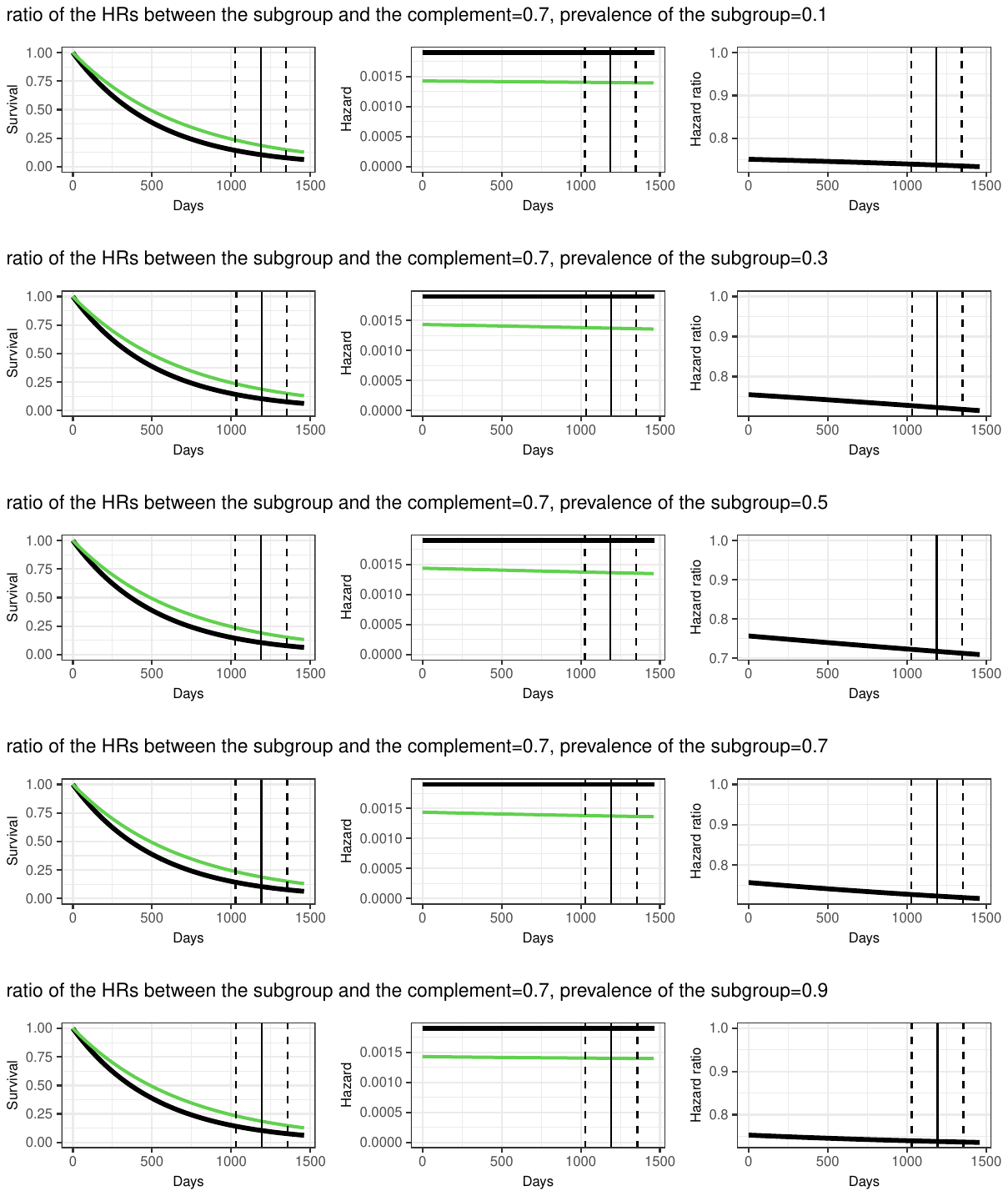}
\end{center}
\caption{Survival functions, hazard functions and hazard ratio functions in the scenarios with disease progression. Each row of the figure represents a scenario with values for the proportion of subjects who progress under treatment and control and the hazard ratio before versus after progression as indicated in the superscripts. The left panel shows survival functions for treatment (green) and control (black). In each row, the center panel shows the corresponding hazard functions. The right panel shoes the hazard ratio function. Additionally, the distribution of the follow-up time across simulation runs is shown in terms of the mean follow-up time, represented by a vertical solid line, and mean +/- 2 standard deviations, represented by vertical dashed lines.}
\label{fig.scenario_progression}
\end{figure}
\end{document}